\documentstyle[psfig]{mn}

\newcommand{\Msun}{M_{\odot}}
\newcommand{\Zsun}{Z_{\odot}}
\newcommand{\spc}{{\rm pc}^{-2}}
\newcommand{\pGyr}{{\rm Gyr}^{-1}}
\newcommand{\CO}{^{12}\mathrm{C}(\alpha,\gamma)^{16}\mathrm{O}}

\newcommand{\etal}{{et~al.}}
\newcommand{\nucl}[2]{\,^{#1}{\rm {#2}}}

\title[Stellar Yields and Chemical Evolution I: the Solar Neighbourhood]
{Stellar Yields and Chemical Evolution I: Abundance
Ratios and Delayed Mixing in the Solar Neighbourhood}

\author[D.~Thomas, L.~Greggio and R.~Bender]
{D.~Thomas,$^{1}$\thanks{daniel@usm.uni-muenchen.de}
L.~Greggio$^{1,2}$\thanks{greggio@usm.uni-muenchen.de} and 
R.~Bender$^{1}$\thanks{bender@usm.uni-muenchen.de}\\
$^1$Universit\"ats-Sternwarte M\"unchen, Scheinerstr.~1, D-81679 M\"unchen,
Germany\\
$^2$Dipartimento di Astronomia, Universit\`{a} di Bologna, I40100 Bologna,
Italy}

\begin{document}

\maketitle
\begin{abstract}
We analyse two recent computations of type II supernova nucleosynthesis by
Woosley \& Weaver (1995, hereafter WW95) and Thielemann, Nomoto, \& Hashimoto 
(1996, hereafter TNH96),
focusing on the ability to reproduce the observed [Mg/Fe]-ratios in various 
galaxy types. We show that the yields of oxygen and total metallicity
are in good agreement. However, TNH96-models produce more magnesium in the
intermediate and less iron in the upper mass range of type II supernovae
than WW95-models.
To investigate the significance of these discrepancies for chemical evolution,
we calculate {\em Simple Stellar Population}-yields for both sets of models
and different IMF slopes. 
We conclude that the Mg-yields of WW95 do not suffice to
explain the [Mg/Fe] overabundance neither in giant elliptical galaxies and
bulges nor in metal-poor stars in the solar neighbourhood and the galactic
halo.
Calculating the chemical evolution in the solar neighbourhood according to
the standard infall-model (e.g.~Matteucci \& Greggio 1986; Timmes, Woosley, \&
Weaver 1995; Yoshii, Tsujimoto, \& Nomoto 1995)\nocite{MG86,TWW95,YTN96}, 
we find that using WW95 and TNH96 nucleosynthesis, the solar
magnesium abundance is underestimated by 29 and 7 per cent, respectively.

We include the relaxation of the instantaneous {\em mixing}
approximation in chemical evolution models by splitting the gas component
into two different phases. In additional simulations of the chemical
evolution in the solar
neighbourhood, we discuss various timescales for the mixing of the stellar
ejecta with the interstellar medium. We find that a delay of the order of
$10^{8}$ years leads to a better fit of the observational data in the
[Mg/Fe]-[Fe/H] diagram without destroying the agreement with solar element
abundances and the age-metallicity relation.
\end{abstract}


\section{Introduction}
\label{sec:intro}
Understanding the formation and evolution of galaxies is difficult because
it involves many different processes that are coupled in a complex way.
In order to simulate a scenario properly, one
has to consider the dynamics of stars, gas and dark matter as well as star
formation, the interaction with the interstellar medium (ISM), and chemical
enrichment \cite{HB90}. 
Since this problem includes many unknown parameters, it is
more effective to decouple various processes and to closely inspect
observational constraints which are relevant to the different parameters.

Chemical evolution models constrain star formation histories, supernova
rates, and abundances in the ISM, in the stars, and in
the intracluster medium (ICM). Thus, trying to reproduce element abundances in
chemical simulations already puts significant constraints on galaxy formation
without considering complicated dynamical aspects.  Hence, in the present stage 
it is more effective to decouple the dynamical and the chemical approach.

\smallskip
Different timescales for the duration of the star forming phase cause
different abundance ratios in the stars and in the ISM. In short phases of
star formation short-living, massive stars govern the enrichment of the
ISM. Thus, in those formation scenarios the abundance ratios reflect the
type II supernova (SN~II) production (Hashimoto, Iwamoto, \& Nomoto 1993).
\nocite{HIN93}

In order to obtain the various element abundances in the solar neighbourhood, 
a modest and continuous star formation rate (SFR) is necessary \cite{Metal89}.
In the standard models (e.g.\ Matteucci \& Greggio 1986; Timmes, Woosley, \&
Weaver 1995; Tsujimoto \etal\ 1995)\nocite{MG86,TWW95,Tetal95}, 
this kind of star formation 
history is obtained with infall of primordial gas on a timescale of 
several Gyr and an SFR 
which depends on the gas density via the Schmidt-law
\cite{S59,S63}. In this scenario, the disk of the Galaxy is assumed to form out
of slowly accreting gas. Since star formation occurs over a long timescale of
$10^{10}$ yr, 
the chemical evolution is noticeably influenced by SN~Ia.

In bulges and elliptical galaxies, [$\alpha$/Fe]-ratios seem to be enhanced with 
respect to solar abundances (Peletier 1989; Worthey, Faber, \& Gonzalez 1992; 
Davies, Sadler, \& Peletier 1993; McWilliam \& Rich 1994).
\nocite{P89,MW94,WFG92,DSP93}
Since the $\alpha$-elements are mainly produced in
massive stars experiencing SN~II explosions
\cite{W86}, and iron is substantially contributed by type Ia supernovae
(SN~Ia),
the chemical history of the light dominating component of the
stellar population in 
bulges and ellipticals must be dominated by massive stars. In chemical
evolution models, this can be realized by either \cite{WFG92,M94}
\begin{enumerate}
\item a flat IMF or
\item a short phase of star formation or
\item a low fraction of close binary systems experiencing SN~Ia.
\end{enumerate}
Chemical evolution models have to constrain and quantify these different 
possibilities.

Although in the pure chemical approach the number of
input parameters is already reduced, there are still plenty of uncertainties
in the calculations. Typical input parameters are the shape and slope of the 
initial mass function (IMF), 
the SFR, the infall rate and the fraction
of close binary systems producing iron via type~Ia supernovae. However, 
chemical evolution is also very sensitive to the adopted 
stellar yields, especially of SN~II 
(see also Gibson 1997).\nocite{G97} Thus, besides the parameters above,
chemical evolution models should always take into account different stellar 
nucleosynthesis prescriptions, which are strongly affected by uncertainties of
stellar evolution models (Thielemann, Nomoto, \& Hashimoto 1996).

\smallskip
In this paper, we compare two recently published nucleosynthesis
calculations for SN~II by:
\begin{enumerate}
\item Woosley \& Weaver \shortcite{WW95}, hereafter WW95
\item Thielemann \etal\ \shortcite{TNH96} and Nomoto \etal\
\shortcite{Netal97}\footnote{In this paper, the results of Thielemann \etal\
(1996) are extended on a larger mass range.}, hereafter TNH96.
\end{enumerate}
We focus on the question, if the considered sets of models are able to
explain an important observed feature of galaxy formation: the [Mg/Fe]
overabundance. 
There is a broad consensus that metal-poor halo stars in
our Galaxy have magnesium enhanced abundance ratios (Gratton \& Sneden 1988;
Magain 1989; Edvardsson \etal\ 1993; Axer, Fuhrmann, \& Gehren 1994; Axer,
Fuhrmann, \& Gehren 1995; Fuhrmann, Axer, \& Gehren 1995).
\nocite{GS88,M89,EAGLNT93,AFG94,AFG95,FAG95}
The exact value of the
enhancement is still debatable, but it seems to converge to $0.3-0.4$ dex 
\cite{TB93,G95}. This observation can be easily understood taking into account
that metal-poor stars form in the early stages of the galaxy formation, 
when the enrichment due to SN~II is dominating chemical evolution. However, 
we will show that there are still unresolved problems caused by uncertainties 
in stellar nucleosynthesis.

As already noted, in elliptical galaxies there are strong indications from 
spectra in the visual 
light that there is a magnesium overabundance of at least 0.2 dex in nuclei of
these galaxies 
(Worthey \etal\ 1992; Weiss, Peletier, \& Matteucci 1995) \nocite{WFG92,WPM95}. 
However, while the halo in the solar
neighbourhood has low metallicities ($-3\leq$[Fe/H]$\leq -1$), the stars
which dominate the visual light in the nuclei of ellipticals have solar or
super-solar $Z$ \cite{Gre97}. Therefore, the [Mg/Fe] overabundance is 
realized at both low and high $Z$ in two considerably different systems.
While a detailed inspection of element abundances in elliptical galaxies
will be the subject of a forthcoming paper,
in this work we concentrate on the chemical evolution of the solar
neighbourhood using WW95 and TNH96 SN~II yields.

In order to calibrate our code, we use the same approach as in the most common 
chemical evolution models (Matteucci \& Greggio 1986; 
Timmes, Woosley, \& Weaver 1995)\nocite{TWW95} for the solar neighbourhood,
performing the calculation for both sets of nucleosynthesis prescriptions.
Taking finite stellar lifetimes into account,
the classical numerical models relax the instantaneous {\em recycling} 
approximation, but usually assume the stellar ejecta to mix 
instantaneously with the ISM \cite{T80}. Only few
attempts have been made in the literature to relax the instantaneous {\em mixing}
approximation in numerical models of chemical evolution (see
discussion in Timmes \etal\ 1995).\nocite{TWW95} We present a modification
of the basic equations \cite{T80} splitting the gaseous mass into two
different phases, one including the stellar ejecta and the second being cool
and well mixed. The mixing process is characterized by a
gas flow from the first, inactive to the active, star forming gas phase.
We additionally present the results of simulations considering this
modification.

\smallskip
In section~\ref{sec:general} we summarize the most important aspects of chemical
evolution, while in section~\ref{sec:yields} we discuss 
stellar yields from SN~II explosions, comparing WW95 and TNH96.
We analyse their influence on chemical evolution by calculating the
SSP-yields of various elements in section~\ref{sec:ssp}.
In section~\ref{sec:solar} we present our model for the chemical evolution
in the solar neighbourhood. In the conclusion we summarize the main results.


\section{Generalities on chemical evolution}
\label{sec:general}

\subsection{The basic equations}
Non-primordial elements develop in a cycle of birth and death of
stars. These form out of the ISM, process elements, and eject them
during the
late stages of their evolution in the form of stellar winds, planetary
nebulae (PN), or supernovae (SN~Ia and SN~II), depending on their main sequence
mass $m$. 
The formation of stars and the re-ejection of gas can be described by the
following phenomenological equations \cite{T80}:
\begin{eqnarray}
dM_{\rm tot}/dt &=& f \\
dM_s/dt &=& \psi -E \label{eq:stars} \\
dM_g/dt &=& -\psi + E + f \label{eq:gas}
\end{eqnarray}
The total baryonic mass $M_{\rm tot}=M_s+M_g$ is governed by infall or outflow
$f$ of material, being either primordial or enriched gas.
The total stellar mass $M_s$ is {\em increasing} according to the SFR 
$\psi$
and {\em decreasing} due to re-ejection $E$ of gas. The total gaseous mass
$M_g$ behaves exactly contrary to $M_s$ with 
the additional component of in-falling or out-flowing gas $f$. 

The ejection rate $E$ is obtained by integrating the ejected mass fraction
$(1-w_m)$,
folded with the SFR and the normalized IMF $\phi$, from the turnoff mass $m_t$ 
to the maximum stellar mass $m_{\rm max}$. 
Here, $w_m$ denotes the mass fraction of the remnant.
\begin{equation}
E(t)=\int_{m_t}^{m_{\rm max}} (1-w_m)\,\psi(t-\tau_m)\,\phi(m)\;dm 
\label{eq:ejection}
\end{equation}
The quantity $\tau_m$ is the stellar
lifetime of a star with mass $m$. In the instantaneous {\em recycling}
approximation, $\tau_m$ is assumed to be negligible in comparison to the time
$t$. This approximation is relaxed in numerical simulations as well as in our
evolution code. This must not be confused with the instantaneous
{\em mixing} approximation (IMA) which is assumed in most chemical evolution
models \cite{MG86,TWW95,Tetal95,PT95}. 
In section~\ref{sec:solar}, we also relax this assumption and take a delay in the
mixing of the stellar ejecta into account.

\smallskip
Parallel to equation~\ref{eq:gas}, the mass production 
of the element $i$ in the ISM ($X_iM_g$) is expressed in the equation below: 
\begin{equation}
d(X_iM_g)/dt=-X_i\,\psi +E_{i}+X_{i,f}\,f \label{eq:gasmetal}
\end{equation}
Here, $X_i$ is the abundance of element $i$ in the ISM, $X_{i,f}$ is
the abundance of element $i$ in the in-falling or out-flowing gas.
The element ejection rate $E_{i}$ is obtained by integrating the ejected mass
fraction $Q_{im}$ (including both initial abundance and newly produced
material) of the element $i$, 
again folded with the SFR and IMF over the appropriate mass range, equivalent to
equation~\ref{eq:ejection}:
\begin{equation}
E_{i}(t)=\int_{m_t}^{m_{\rm max}} Q_{im}\;\psi(t-\tau_m)\,\phi(m)\;dm 
\label{eq:metal_eject}
\end{equation}
Equations~\ref{eq:gas} and~\ref{eq:gasmetal} can be combined to describe 
the progression of the abundance $X_i$ of element $i$ in the ISM.
\begin{equation}
M_g\,dX_i/dt = E_{i}-X_i\,E + (X_{i,f}-X_i)\:f
\label{eq:abundance}
\end{equation}

The key value in this equation is the stellar yield $Q_{im}$ hidden in the
element ejection rate $E_i$.We neglect stellar winds during the evolution,
and assume that the stars enrich the ISM at the time when they die. 
Depending on their initial
mass, they either experience a SN~II explosion ($m>8~\Msun$) or 
become a white dwarf blowing off their envelopes 
($m<8~\Msun$). Elements heavier than oxygen are mainly processed in 
supernovae. A substantial fraction of iron is contributed by SN~Ia.
Adopting the description of the
supernova rates from Greggio \& Renzini \shortcite{GR83}, 
the element ejection rate integrated over the total mass range can be
described with the following equation as in Matteucci \& Greggio (1986) and
Timmes \etal\ (1995):
\begin{eqnarray}
E_i(t) &=& \int_{16}^{m_{\rm max}} Q_{im}^{\rm SNII}\psi(t-\tau_m)\phi(m)\: dm \nonumber\\
&+& (1-A)\: \int_8^{16} Q_{im}^{\rm SNII}\psi(t-\tau_m)\phi(m)\: dm 
\nonumber\\
&+& (1-A)\int_{3}^{8} Q_{im}^{\rm PN}\psi(t-\tau_m)\phi(m)\: dm\nonumber\\
&+& A\:\int_3^{16} \phi(m)\, dm \int_{\mu_{\rm inf}}^{0.5} 24\mu^2 
Q_{im}^{\rm SNIa}\psi(t-\tau_{\mu m})\, d\mu \nonumber\\
&+& \int_{1}^{3} Q_{im}^{\rm PN}\psi(t-\tau_m)\phi(m)\: dm
\label{eq:all_enrich}
\end{eqnarray}
In this equation, the enrichment due to stars in the mass range $3-16~\Msun$
is splitted into a
contribution by type II supernovae ($Q^{\rm SNII}$) plus planetary nebulae
($Q^{\rm PN}$), due to single stars, and type Ia supernovae
($Q^{\rm SNIa}$), assumed to be the end product of close binary evolution.
In the formulation of Greggio \& Renzini (1983) $\mu$ is the ratio between the
mass of the secondary and the total mass $m$ of the system. The maximum
fraction of the secondary is 0.5 by definition, while the minimum mass
$\mu_{\rm inf}$ is dependent on the turnoff as defined in the following 
equation \cite{GR83}:
\begin{equation}
\mu_{\rm inf}\equiv {\rm MAX}\;[m_t/m,(m-8)/m]
\end{equation}
One has to integrate the distribution of the secondary component
$f(\mu)\sim \mu^2$, folded with the yield (independent of the mass of the
system) and the SFR over the appropriate mass range.
The clock of the SN~Ia explosion is given by the
lifetime of the secondary $\tau_{\mu m}$, whose mass can be as low as
$0.8~\Msun$ \cite{GR83}. Thus, the enrichment due to SN~Ia
is substantially delayed with respect to SN~II.
The degree of influence by SN~Ia highly depends
on the fraction $A$ of close binaries, which is a free 
parameter in chemical evolution. Greggio \& Renzini \shortcite{GR83}
calibrate $A$ on the ratio between the
current type II to type Ia supernova rates in the Galaxy.

It should be noticed that equation~\ref{eq:all_enrich} is not completely
consistent, since $m$ refers to the mass of single and binary stars.
Nevertheless, as long as the parameter $A$ is small (as in our case) type Ia
events can be regarded as a small perturbation. Hence, equation~\ref{eq:all_enrich}
is an acceptable approximation, and it allows us to describe the delayed
release of iron from SN~Ia.

In these terms, the rates of type II and Ia supernovae can be described by
the following equations:
\begin{eqnarray}
R_{\rm II}&=&\int_{16}^{m_{\rm max}} \psi(t-\tau_m)\frac{\phi(m)}{m}\: dm \nonumber\\
          &+&(1-A)\: \int_{8}^{16}\psi(t-\tau_m)\frac{\phi(m)}{m}\: dm
\label{eq:SNII}
\end{eqnarray}
\begin{equation}
R_{\rm Ia}=A\:\int_3^{16} \frac{\phi(m)}{m}\, dm \int_{\mu_{\rm inf}}^{0.5} 24\mu^2 
\psi(t-\tau_{\mu m})\, d\mu 
\label{eq:SNIa}
\end{equation}

\subsection{Parameter constraints}
In spite of the several unknown parameters (IMF slope $x$, SFR as a function
of time, stellar
yields), the following arguments show how they can be constrained by
different observational information, step by step.
This is useful for the
interpretation of the results from chemical evolution calculations. 

Assuming the
instantaneous {\em recycling} approximation, one can take $\psi$ out of the
integral in eqn.~\ref{eq:ejection}.
For large $t$, the residual
integral expressing the returned fraction $R_x$ depends strongly
on the IMF slope $x$
but only marginally on turnoff mass (and then time). Equation~\ref{eq:stars} can
then be written as:
\begin{equation}
dM_s/dt\approx\psi(1-R_x) \label{eq:stars_approx}
\end{equation}
The integrated solution demonstrates that the final total mass of stars
depends on the {\em time-average} star formation rate $\bar{\psi}$ and the IMF
slope.
\begin{equation}
M_s\approx M_{s,0} + (1-R_x)\; \int_{t_0}^t \psi(t')\: dt'\ ,
\label{eq:stars_estimate}
\end{equation}
where the subscript zero refers to the initial conditions.
Similar considerations show that the final mass of gas in the ISM depends on a
mean SFR $\bar{\psi}$, a mean infall rate $\bar{f}$ and the IMF slope.
\begin{equation}
M_g\approx M_{g,0} - (1-R_x)\; \int_{t_0}^t \psi(t')\: dt' +
\int_{t_0}^t f(t')\: dt'
\end{equation}

\smallskip
Now we need an approximation to pin down $x$. For this purpose, we
consider the element abundances at the time $t$. 
The above approximation applied
to equation~\ref{eq:gasmetal} leads to
\begin{eqnarray}
X_iM_g &\approx& X_{i,0}M_{g,0} - (X_i-X_{i,0})\: \bar{\psi}\: (t-t_0)
\nonumber\\
&+& R_{ix}\: \bar{\psi}\: (t-t_0) +\ \left(X_{i,f}-X_{i,f,0}\right)\: \bar{f}\:
(t-t_0)
\label{eq:gasmetals_estimate}
\end{eqnarray}
with $R_{ix}$
as the returned mass fraction of element $i$ mostly dependent on the IMF slope
and the stellar yield. If we consider an element whose stellar yield is
well known, we get a constraint on $x$ (with known abundance of the in-falling
gas), for a given minimum mass of the stellar population $m_{\rm min}$.

To summarize, the stellar and gaseous mass at the current epoch and the
abundance of a specific element whose yield is relatively certain 
constrain $\bar{\psi},\bar{f}$ and $x$. We can then draw
conclusions on the nucleosynthesis of other elements, whose stellar yields are
relatively uncertain. In other words, from chemical evolution of galaxies one can get a
constraint on the stellar evolution models. 
This is a convincing example how tight these
two disciplines are coupled. 
Since the approximation $\tau_m\ll t$ is especially valid for elements produced
mainly by type II supernovae,
we will use this strategy in section~\ref{sec:solar}
to fix the IMF slope with the element oxygen and constrain the necessary
magnesium yield in SN~II.

\subsection{The initial mass function}
The IMF is as usual assumed to be a declining function of mass, according to
a power law: $\phi\sim m^{-x}$. Since the IMF is usually normalized 
{\em by mass}, the actual amount of mass created in one generation of stars is 
controlled by the SFR $\psi$.
\begin{equation}
\int_{m_{\rm min}}^{m_{\rm max}} \phi(m)\: dm = 1
\end{equation}
In these terms, the slope $x=1.35$ corresponds to the Salpeter-value \cite{S55}.

To avoid uncertain extrapolations of the stellar yields to the high mass
end, we have adopted $m_{\rm max}=40~\Msun$ which is the maximum mass for
which WW95-models are computed. TNH96 do give the yields for a
$70~\Msun$-star. For the comparison between the two sets of models we keep
$m_{\rm max}$ fixed at $40~\Msun$. The effect of adopting $m_{\rm max}=70~\Msun$
is explored in section~\ref{sec:cutoff} and~\ref{par:cutoff}.

The lower cutoff of one generation of stars is assumed to be
$m_{\rm min}=0.1~\Msun$. The higher the minimum mass, the larger is the
fraction of massive stars, thus more metals are produced.
Abundance ratios, however, are not affected by the choice of $m_{\rm min}$.

Alternative formulations of the IMF with different slopes at different mass
ranges exist in the literature
(e.g.\ Scalo 1986; Kroupa, Tout, \& Gilmore 1995; Gould, Bahcall, \& Flynn
1997).\nocite{S86,KTG93,Getal97}
However, in order to keep the number of free parameters low,
we have decided to fix $m_{\rm min}=0.1~\Msun$ and use one specific slope
$x$ for the whole mass range. The value of $x$, instead, is treated as a free 
parameter.


\section{Stellar yields and nucleosynthesis}
\label{sec:yields}
\subsection{PN and SN~Ia}
In our calculations, we use the results in Renzini \& Voli (1981) for the
enrichment due to intermediate mass single stars ($1\leq m\leq 8~\Msun$). In
particular we select the models with $\alpha=1.5$, $\eta=0.33$. 

Type Ia SNe are assumed to occur in close binary systems \cite{WI73}. In
this model, the explosion is caused by a carbon-deflagration of the material
accreting degenerate white dwarf (Hansen \& Wheeler 1969; Nomoto 1980;
Weaver \& Woosley 1980; Nomoto 1981).\nocite{HW69,N80a,N80b,WW80,N81}
We adopt the results of the nucleosynthesis
from the classical W7-model by Nomoto, Thielemann, \& Yokoi
(1984).\nocite{NTY84}

Low mass stars, in the range $1~\Msun$ to
$8~\Msun$, do not contribute to the enrichment of O, Mg and Fe \cite{RV81}.
Type Ia supernovae produce significantly more iron than oxygen or magnesium
as can be seen in table~\ref{tab:SNIa-yields}. One
can see that $^{56}$Fe is clearly dominating the ejecta. It follows that
SN~II must be the main contributor to the $\alpha$-elements enrichment.
\begin{table}
\caption{The most abundant elements ejected in a type Ia supernova. 
In the calculations, Nomoto \etal\ (1984) assume
accreting white dwarfs in close binary systems to be the progenitors of
SN~Ia events. The data refer to the W7 model, the values are given in $\Msun$.
The numbers show that the ejecta of SN~Ia are clearly dominated
by $\nucl{56}{Fe}$.}
\begin{tabular}{lrlrlr}
\hline
$^{12}$C  & 3.2e-2 & $^{28}$Si & 1.6e-1 & $^{56}$Fe & 6.1e-1\\
$^{16}$O  & 1.4e-1 & $^{32}$S  & 8.2e-2 & $^{57}$Fe & 1.1e-2\\
$^{20}$Ne & 1.1e-2 & $^{36}$Ar & 2.2e-2 & $^{58}$Ni & 6.1e-2\\  
$^{24}$Mg & 2.3e-2 & $^{40}$Ca & 4.1e-2 & $^{60}$Ni & 1.1e-2\\
\hline
\end{tabular}
\label{tab:SNIa-yields}
\nocite{NTY84}
\end{table}

\subsection{SN~II}
As mentioned in the introduction, we use two sets of models for the
enrichment due to type II supernova explosions: WW95 and TNH96.

The calculation of the SN~II yields is affected by many uncertainties (see
WW95, TNH96 and references therein).
Elements lighter than iron like carbon, oxygen, and magnesium are mainly
produced during the evolutionary stages of the star {\em before} the
explosion \cite{WeWo95}. Thus, their abundances in the SN~II-ejecta are highly
dependent on stellar evolution, especially on the $\CO$-rate during He-burning 
and the treatment of convection. Both, a higher $\CO$-rate and 
the inclusion of semi-convection lead to a smaller production of carbon and
carbon-burning products (TNH96).

The iron produced in hydrostatic silicon burning during the pre-supernova
evolution forms the core of the star, which represents the minimum mass of the
remnant. Depending on the position of the mass-cut and the fraction of mass
falling back, the remnant mass can be higher \cite{NH88,HNTT93,WW93}.
The total amount of iron in the ejecta is exclusively
produced during the explosion. More precisely, most of the explosively generated
$^{56}$Ni decays to $^{56}$Fe. Thus, the theoretical iron yield of a SN~II 
does not directly depend on parameters of stellar evolution,
but on the simulation of the explosion itself.

Table~\ref{tab:WW_TNH} shows the important differences between the two
sets.
\begin{table*}
\begin{minipage}{15cm}
\caption{The main differences in the SN~II nucleosynthesis prescriptions of
WW95 and TNH96. Models B and C in WW95 refer to enhanced explosion energies
in high mass stars by a factor of 1.5 and 2, respectively. TNH96 do not
specify different models, but also enhance the explosion energy in high mass
stars by a factor of 1.5. Differences in stellar evolution ($\CO$-rate,
convection theory) mainly affect the nucleosynthesis of intermediate
elements lighter than iron. The yield of iron itself is highly dependent on
the explosion.}
\begin{tabular}{lll}
\hline
& WW95 & TNH96 
\\\\
$\CO$      & $1.7\times$ Caughlan \& Fowler (1988), & Caughlan \etal\ (1985)
\\
           & 74 per cent of TNH96                   &
\\
convection & Ledoux criterion,                      & Schwarzschild criterion, 
\\
           & modification for semi-convection       & convective shells have greater extend 
\\
explosion energy & $1.2\times 10^{51}$ erg (model A) & $1.0\times 10^{51}$ erg
\\
           & model B: $E_{\rm B}\approx 1.5\times E_{\rm A}$ for $m\geq 30~\Msun$ & $E=1.5\times 10^{51}$ erg for $m\geq 25~\Msun$
\\
           & model C: $E_{\rm C}\approx 2\times E_{\rm A}$ for $m\geq 35~\Msun$ &
\\
explosion mechanism & piston situated at the $Y_e$ discontinuity & deposition of energy 
\\
neutrinos  & nucleosynthesis caused by the       & neutrino process {\em not} included
\\
           & flood of neu\-trinos                &
\\
mass grid  & 11, 12, 13, 15, 18, 19, 20, 22,     & 13, 15, 18, 20, 25, 40, 70 $\Msun$
\\
           & 25, 30, 35, 40 $\Msun$              &
\\
initial metallicity & grid of 5 different $Z_{\rm in}$ & only solar $Z_{\rm in}$ 
\\
stellar evolution & entire stars & helium cores 
\\ \hline
\end{tabular}
\label{tab:WW_TNH}
\end{minipage}
\nocite{CF88,CFHZ85}
\end{table*}
WW95 specify models A, B and C. In model B, the explosion energies are
enhanced by a factor $\sim 1.5$ in stars with $m\geq 30~\Msun$,
in model C by a factor $\sim 2$ in stars with $m\geq 35~\Msun$, both with
respect to model A. 
TNH96 enhance the explosion energy for $m\geq 25~\Msun$ by a factor of 1.5
with respect to the lower masses, as well. 
Hence, their models correspond best to model B in WW95. In the following, 
if not otherwise specified, the considered WW95-models are model B.

We discuss the differences in the yields of
H, He, O, Mg, Fe and total ejected metals $Z_{\rm ej}$
as functions of the main sequence mass of the star $m$ ($\Msun$).
TNH96 evolve helium cores of mass $m_{\alpha}$, adopting the relation between
$m$ and $m_{\alpha}$ from Sugimoto \& Nomoto \shortcite{SN80}.
The total ejected mass of a certain element is then
given by the calculated yield from the evolution of $m_{\alpha}$ plus the
original element abundance in the envelope $m-m_{\alpha}$. Since
TNH96 consider solar initial metallicity, for the discussion of the yields we
assume the element abundances in the envelope to be solar. We use the
solar element abundances from Anders \& Grevesse \shortcite{AG89}, meteoritic
values. It should be mentioned that the $70~\Msun$-star of the TNH96-results 
is not shown in the plots, but the relative yields are given in the captions.

\subsubsection{Ejected mass and hydrogen}
\begin{figure}
\psfig{figure=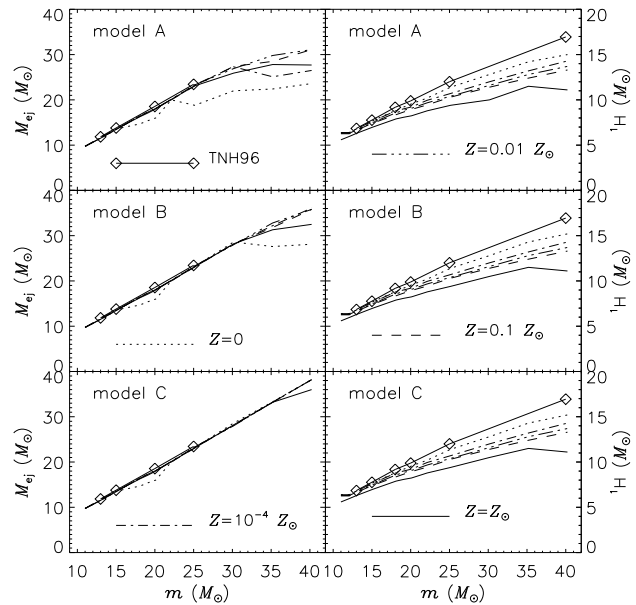,width=\linewidth}
\caption
{Total ejected mass (left panel) and hydrogen yield (right panel) of SN~II as a 
function of initial stellar mass ($\Msun$). 
In each panel, one of the different linestyles is defined, indicating the 
five different initial metallicities assumed in WW95. The diamonds refer to the 
results of TNH96
($\Zsun$). The second and third row show the results for enhanced explosion 
energy in high mass stars in the calculations of WW95. The yields include
initial and newly synthesized material. The hydrogen yield in the
WW95-models shows a clear dependence on metallicity for high mass stars.}
\label{eject_hydro}
\end{figure}
The left panels in figure~\ref{eject_hydro} show the total ejected masses. 
Models A 
(WW95) are characterized by the fall-back of envelope material in the high 
mass range, an effect less pronounced in models B and virtually absent in
models C. Except for the case $Z=0$, the dependence on metallicity seems 
unimportant. The ejected masses in the WW95 and TNH96 models are very similar.

Conversely, the hydrogen yield (figure~\ref{eject_hydro}, right panels)
is clearly dependent on the initial metallicity of
the star, especially at the high mass end. Furthermore, for $m\ga 20~\Msun$
the H-yield given by TNH96 is larger than that in WW95-models 

Both prescriptions basically agree in the value of
$m_{\rm H}+m_{\rm He}$. Table~\ref{tab_helium} shows that the higher 
value for $m_{\rm H}$ corresponds to a lower helium yield in TNH96.
\begin{table}
\caption{SN~II $\nucl{4}{He}$ yields according to WW95 ($\Zsun$, model B) and 
TNH96. The numbers are given in $\Msun$. Since Nomoto et~al.\ (1997)
do not give He-yields, the considered (smaller) mass grid is taken from
Thielemann et~al.\ (1996).}
\begin{tabular}{lcccc}
\hline
      & $13~\Msun$  & $15~\Msun$  & $20~\Msun$  & $25~\Msun$ \\
WW95  & 4.51        & 5.24        & 6.72        & 8.64  \\
TNH96 & 4.13        & 4.86        & 5.95        & 6.63     \\ \hline
\end{tabular}
\label{tab_helium}
\end{table}

The difference in hydrogen (and then helium) yields
comes from two causes: a different $m_{\alpha}$-$m$ relation at He
ignition and the fact that TNH96 neglect the H-shell burning occurring
after He ignition.
In this respect we notice that in WW95 models, the $m_{\alpha}$ of
a $25~\Msun$ star is $9.21~\Msun$, $1.21~\Msun$ larger than that
adopted by TNH96 for the same initial mass, on the basis of the
$m_{\alpha}$-$m$ relation by Sugimoto \& Nomoto \shortcite{SN80}.
The He-yield of this star is $2~\Msun$ larger than in TNH96 models, reflecting
the He-production due to the H-burning shell. A fair comparison between the
predictions of the two sets of models should be done at constant
$m_{\alpha}$. However, since we lack the $m_{\alpha}$-values for WW95 models
for masses other than $25~\Msun$, we proceed comparing the elements
production for the same initial mass. 

\subsubsection{Oxygen and metallicity}
\label{section_oxy_z}
\begin{figure}
\psfig{figure=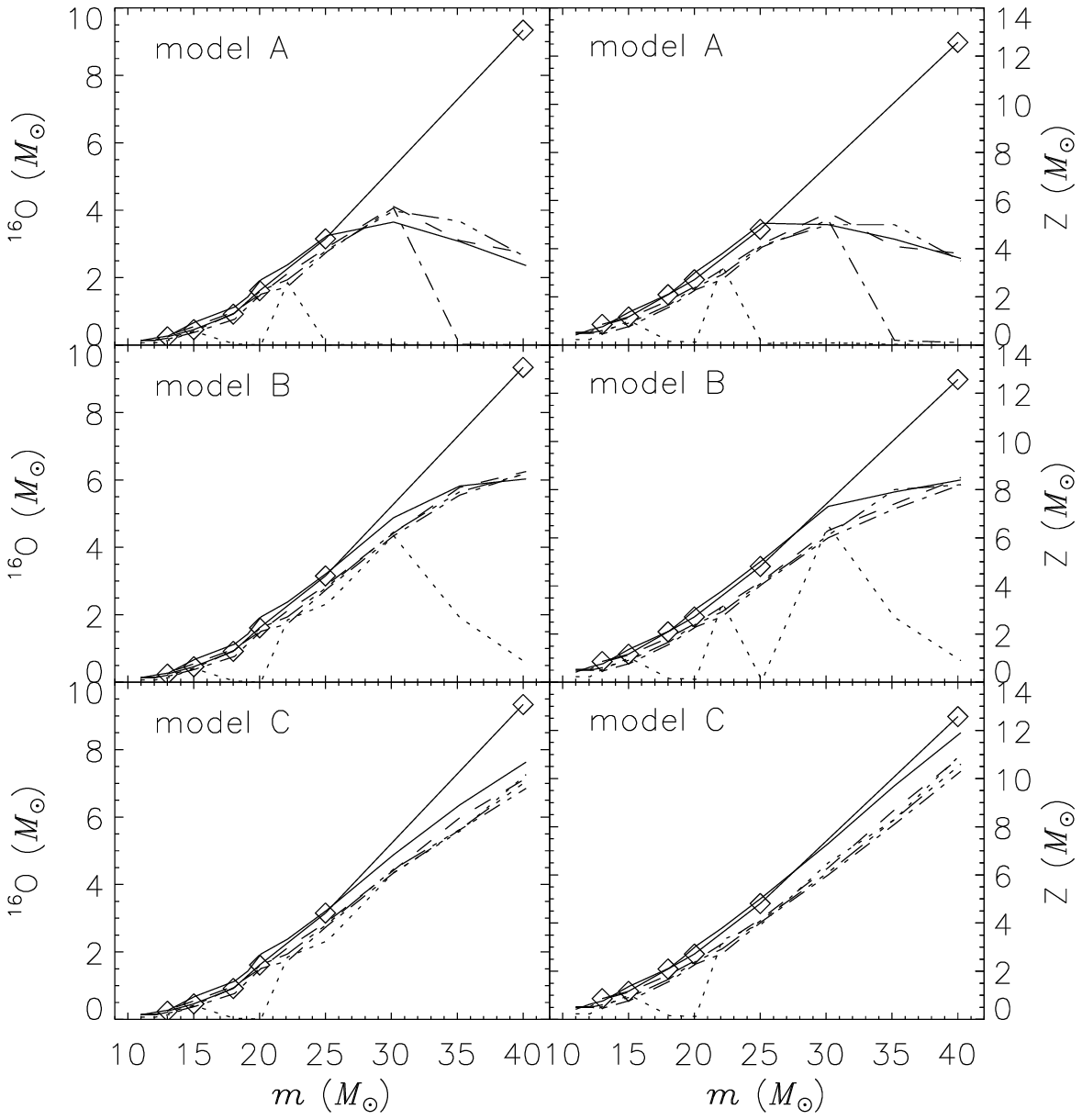,width=\linewidth}
\caption
{Oxygen yield (left panel) and ejected metallicity (right panel)
of SN~II as a function of
initial stellar
mass ($\Msun$). The different linestyles and symbols are explained in
figure~\ref{eject_hydro}. The yields include initial and newly synthesized
material. The figure demonstrates that oxygen is clearly dominating the
total metallicity of the ejecta. The dependence on initial metallicity of
the star seems negligible (except for $Z_{\rm in}=0$). The oxygen yield of
the $70~\Msun$-star according to TNH96 is $\sim 22~\Msun$.}
\label{oxy_z}
\end{figure}

The yield of oxygen and total ejected mass of all elements heavier than helium
($Z$) are plotted in figure~\ref{oxy_z}. The figure shows that
$Z$ is clearly dominated by oxygen. Both depend only weakly on
initial metallicity except for the $Z_{\rm in}=0$ case. The results of WW95 and
TNH96 are similar, except that TNH96 produce more oxygen in the higher mass
range. This can be understood in terms of the higher $\CO$-rate in the TNH96
models. It is worth noting that the large difference in the yields of high
mass stars may also result from the fact that WW95 consider fall back of
material, whereas TNH96 do not. The O-yield of TNH96 increases rapidly with
mass, the WW95-yields, instead, seem to saturate. This discrepancy already
indicates that there is a huge uncertainty concerning the stellar yields of
high mass stars ($m>40~\Msun$).

At the lower mass range ($m\leq 20~\Msun$), WW95 yields tend to be slightly 
larger for the same metallicity ($\Zsun$), possibly because of the larger 
$m_{\alpha}$.
However, the similarity of the results of the two sets of models suggests
that the uncertainty in the oxygen yield from SN~II is small.

\subsubsection{Magnesium and iron}
\begin{figure}
\psfig{figure=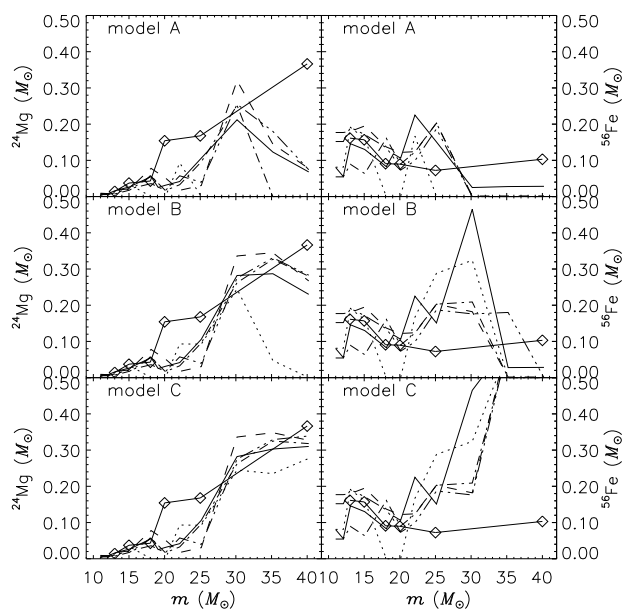,width=\linewidth}
\caption
{Magnesium yield (left panel) and iron yield (right panel)
of SN~II as a function of initial
stellar mass ($\Msun$).
The different linestyles and symbols are explained in figure~\ref{eject_hydro}.
The yields include initial and newly synthesized material. The dependence of
both, the Mg- and Fe-yields, on initial metallicity is not very clear. TNH96
and WW95 agree very well in the Mg- and Fe-yields for low mass stars.
The magnesium and iron  yields of the $70~\Msun$-star according to TNH96 are
$\sim 0.8~\Msun$ and $\sim 0.1~\Msun$, respectively.}
\label{iron_mag}
\end{figure}
Both, WW95 and TNH96 produce a Mg-yield with a rapid rise in a certain mass
range at $M\approx 18~\Msun$ for TNH96 and at $M\approx 23~\Msun$ for WW95.
As a consequence, in the mass range $18-25~\Msun$, the Mg yield of TNH96 is
larger by about a factor of $3-5$. We will briefly investigate the origin of
this discrepancy, which, as we will show, is very significant in the context
of chemical evolution.

\smallskip
$\nucl{24}{Mg}$ is mainly produced during hydrostatic carbon-burning.
Thus, table~\ref{tab:diffmag} gives the yields of $\nucl{12}{C}, \nucl{16}{O}$
and of the main carbon-burning products
$\nucl{20}{Ne}, \nucl{23}{Na}, \nucl{24}{Mg}$ \cite{AT85}.
$\nucl{24}{Mg}$ is produced in the following reaction \cite{AT85}:
\[ \nucl{12}{C}(\nucl{12}{C},p)\nucl{23}{Na}(p,\gamma)\nucl{24}{Mg}\ . \]
Hence, the model producing more carbon should also produce more magnesium.
Table~\ref{tab:diffmag} shows, that the carbon yields are
systematically higher in WW95 for all stellar masses. This is reasonable when
taking into account
the larger helium cores\footnote{Carbon is a helium-burning product.}
and the lower $\CO$-rate of WW95. However, for the yields of
$\nucl{20}{Ne}$, $\nucl{23}{Na}$ and $\nucl{24}{Mg}$ this is not the case
for all masses.
In general, for low mass stars ($m\leq 18~\Msun$) the yields of the
carbon-burning products $\nucl{20}{Ne}$ and $\nucl{23}{Na}$ are higher in
WW95 models as well. With the exception of the $40~\Msun$-star, the higher
masses exactly inverse this pattern. The yield of $\nucl{24}{Mg}$ behaves
similar, but the effect is much stronger with the largest
discrepancy for the $20~\Msun$-star. WW95 argue that the larger extent
of the convective shells in the TNH96-models (Schwarzschild criterion) is
responsible for the above behaviour.
Since the observations of magnesium overabundance can be better explained
with high Mg-yields in SN~II (see following sections), this could be
interpreted as an argument in favour of the Schwarzschild criterion in
convection theory.
\begin{table*}
\begin{minipage}{12cm}
\caption{SN~II yields of the elements C, O, and the main carbon-burning
products; comparison of WW95 ($\Zsun$, model B) and TNH96. The numbers are
given in units of $\Msun$. The TNH96 numbers consist of the given yield from
helium core evolution plus the initial abundance (solar) 
of the envelope (see text).
In spite of the systematically larger $\nucl{12}{C}$-yield,
WW95 give less $\nucl{24}{Mg}$ in most of the stars, although magnesium is a
carbon-burning product. Due to WW95, this pattern is caused by the different
convection theories (see text).}
\begin{tabular}{rrllllll}
\hline
         &       &$13~\Msun$&$15~\Msun$&$18~\Msun$&$20~\Msun$&$25~\Msun$&$40~\Msun$ 
\\\\
$^{12}$C & WW95  &$1.14(-1)$&$1.61(-1)$&$2.48(-1)$&$2.13(-1)$&$3.22(-1)$&$3.63(-1)$ \\
         & TNH96 &$3.21(-2)$&$1.16(-1)$&$2.04(-1)$&$1.56(-1)$&$2.00(-1)$&$2.21(-1)$ \\
$^{16}$O & WW95  &$2.72(-1)$&$6.80(-1)$& 1.13     & 1.94     & 3.25     & 6.03 \\
         & TNH96 &$2.44(-1)$&$4.60(-1)$&$9.17(-1)$& 1.61     & 3.15     & 9.34 \\
$^{20}$Ne& WW95  &$4.46(-2)$&$1.11(-1)$&$2.77(-1)$&$1.05(-1)$&$3.94(-1)$& 1.24 \\
         & TNH96 &$3.82(-2)$&$3.86(-2)$&$1.82(-1)$&$2.52(-1)$&$6.22(-1)$&$6.97(-1)$ \\
$^{23}$Na& WW95  &$1.08(-3)$&$3.42(-3)$&$9.99(-3)$&$1.53(-3)$&$1.08(-2)$&$3.68(-2)$ \\
         & TNH96 &$1.05(-3)$&$5.20(-4)$&$7.68(-3)$&$1.62(-3)$&$1.87(-2)$&$2.45(-2)$ \\
$^{24}$Mg& WW95  &$1.64(-2)$&$2.67(-2)$&$5.52(-2)$&$3.13(-2)$&$1.06(-1)$&$2.30(-1)$\\
         & TNH96 &$1.42(-2)$&$3.73(-2)$&$4.29(-2)$&$1.54(-1)$&$1.68(-1)$&$3.66(-1)$ \\
\hline
\end{tabular}
\label{tab:diffmag}
\end{minipage}
\end{table*}

Similar to the O-yields, the Mg-yields of WW95 seem to saturate or even
decline for increasing mass above $40~\Msun$, due to re-implosion. According 
to the TNH96 calculations, instead, a huge amount of magnesium is ejected by 
high mass stars. 

\smallskip
Figure~\ref{iron_mag} shows, that the iron yield declines for masses between
$13~\Msun$ and $20~\Msun$ in both sets of models.
The iron yields in the lower mass range are very similar, both
models match the observational constraints at $14~\Msun$ (SN1993J; e.g.~Baron, 
Hauschildt, \& Young 1995; Nomoto, Iwamoto, \& Suzuki 1995) and
$20~\Msun$ (SN1987A; e.g~Arnett \etal\ 1989). 
Table~\ref{tab:ni_obs} shows that both groups can reproduce the
observed $\nucl{56}{Ni}$ of the supernova events, which is dominating the
iron yield. Thus, in the lower mass range, TNH96 and WW95 basically agree in
the Fe-yield.
\begin{table}
\caption{Theoretical and observed ejected $^{56}$Ni ($\Msun$) in the SN~II 
events SN1993J ($14~\Msun$; Arnett \etal\ 1989) and SN1987A
($20~\Msun$; Baron \etal\ 1995; Nomoto \etal\ 1995).
The observational data are compared with the theoretical results of TNH96
and WW95. Both nucleosynthesis prescriptions are in agreement with observation.}
\begin{tabular}{llll}
\hline
$m_{*}$ ($\Msun$) & observation         & WW95 ($\Zsun$)  & TNH96  \\\\
$14\pm 1$         & $0.100\pm 0.02$     & $0.133-0.115$   & $0.153-0.130$\\
$20$              & $0.075\pm 0.01$     & 0.088           & 0.074 \\\hline
\end{tabular}
\label{tab:ni_obs}
\nocite{ABKW89,BHY95,NIS95}
\end{table}

However, WW95 produce significantly more iron than TNH96 in stars of 
$m\geq 25~\Msun$, especially in model B and C. Thus, mainly this mass range
will be responsible for discrepancies in the Fe-yields of the total mass
range of SN~II (SSP-yields, see section~\ref{sec:ssp}).

\subsection{IMF-weighted yields}
For the discussion of chemical evolution, it is more meaningful to
consider stellar yields weighted by the IMF. Normalized on the SN~II yield
of the whole mass range, these values give the relative
contribution of a $1~\Msun$ interval to the total SN~II yield. To show
the role of various mass intervals to the enrichment of a certain element,
we plot the IMF weighted yields of the elements oxygen, magnesium, iron, and
metallicity for different IMF slopes and both nucleosynthesis models WW95
and TNH96. 
In more detail, we plot the following quantity:
\begin{equation}
\frac{dQ_{im}}{dm}= \frac{Q_{im}\times \phi(m)}
{\int_{11}^{40} Q_{im}\times \phi(m)\: dm}
\label{eq:cont}
\end{equation}
The figures are given in the appendix~\ref{ap:cont}.
Summarizing the plots, we obtain the following results:
\begin{enumerate}
\item There is no specific mass range dominating the O- and $Z$-enrichment
significantly for all considered IMF slopes.
\item The IMF weighted Mg-yield is slightly peaked at $30~\Msun$ (WW95) and $20~\Msun$
(TNH96), but again these masses do not dominate the SSP-yield significantly.
\item The Fe-enrichment due to SN~II, instead, is clearly governed by stars of 
$m\leq 20~\Msun$.
\end{enumerate}
Altogether, the plots demonstrate the increasing weight of the higher mass
range with decreasing IMF slope $x$.

\subsection{The ratio [Mg/Fe]}
\begin{figure}
\psfig{figure=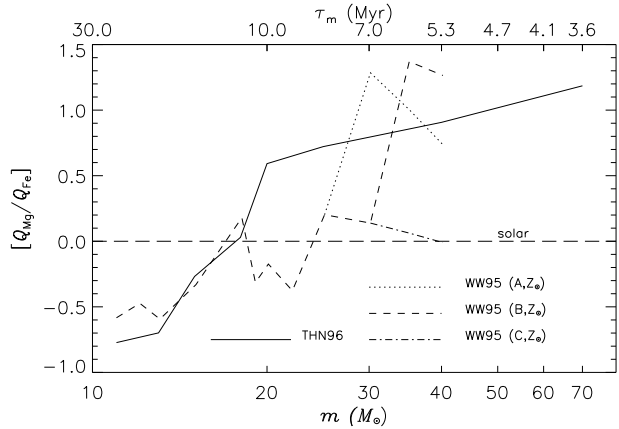,width=\linewidth}
\caption
{Abundance ratio of magnesium to iron in the ejecta of type II SNe
as a function of initial stellar mass. The values are normalized on solar
values and plotted on a logarithmic scale. The WW95-models of solar
initial metallicity are considered. The upper x-axis denotes the lifetime
$\tau_m$ of the star with mass $m$ (Schaller \etal~1992). Mainly stars in the
upper mass range contribute to super-solar [Mg/Fe]-ratios.}
\label{cont_mgfe}
\end{figure}

In figure~\ref{cont_mgfe},
we plot the abundance ratio [Mg/Fe] produced in WW95 ($\Zsun$) and
TNH96 as a function of stellar mass.
The figure shows that the ratio of magnesium to iron is basically increasing
with mass. In the intermediate mass range, the overabundance in TNH96 models
exceeds the results of WW95 significantly. The maximum overabundance in
TNH96 is reached in the most massive star ($m=70~\Msun$), whereas Mg/Fe of the
WW95-models peaks at $m=35\Msun$. 

According to WW95, a magnesium overabundance is only produced in
stars with $m\ga 25~\Msun$ (except for the small peak at $m=18~\Msun$).
Thus, in the first $\sim 10$ Myr, when the turnoff is above $20~\Msun$ (see
upper x-axis in figure~\ref{cont_mgfe}),
these stars will enrich the ISM with highly magnesium overabundant ejecta.
But already 30 Myr after the beginning of star formation, the turnoff of
$10~\Msun$ is reached and the whole SN~II generation of stars is contributing 
to the enrichment. Thus, the key value for the discussion of chemical
evolution is the SSP-yield.


\section{SSP-yields}
\label{sec:ssp}
We calculated SSP-yields of the elements oxygen, magnesium, iron in the mass
range of type II SNe for different IMF slopes and both sets of SN~II
nucleosynthesis. The tables in the appendix~\ref{ap:tables} give the abundances
of the considered elements in the ejecta of SN~II explosions of one SSP
($m_{\rm max}=40~\Msun$).
The basic conclusion for the discussion of the yields are:
\begin{enumerate}
\item The highest [Mg/Fe]-ratio in WW95 is produced in model B assuming an
initial metallicity of $Z=10^{-4}\Zsun$. This ratio is lowest for models C
because of the high iron yield.
\item The second highest value for [Mg/Fe] is produced in the models with
$Z=\Zsun$. The results for $Z=0.01\Zsun$ and $Z=0.1\Zsun$ are in between.
Hence, [Mg/Fe] is neither increasing nor decreasing with initial
metallicity. The high and low WW95 metallicities {\em do not} bracket the
expected SSP-yields as claimed by Gibson, Loewenstein, \& Mushotzky
\shortcite{GLM97}.
\item TNH96 produce systematically higher [Mg/Fe]-ratios than WW95.
\end{enumerate}
For Salpeter-IMF, the magnesium abundance in the SN~II ejecta is 0.13 dex
higher with TNH96-models than with WW95-models. The iron abundance, instead,
is 0.08 dex lower. In total, this leads to a [Mg/Fe]-ratio, which is 0.21
dex higher for TNH96 nucleosynthesis.

\smallskip
We define the {\em time dependent }SSP-yield for element $i$ at time $t$ as
\begin{equation}
Q_{\rm SSP}^i(t) = \frac{\int_{m_t}^{m_{max}} Q_{im}\, \phi(m)\: dm}
                      {\int_{m_t}^{m_{max}} (1-w_m)\, \phi(m)\: dm}
\label{time_SSP}
\end{equation}
This equation describes the abundance of element $i$ in the ejecta of one
generation of stars of one single metallicity at the time $t$. With
progression of time, the turnoff mass decreases and $Q_{\rm SSP}^i(t)$ converges
to the standard SSP-yield, integrated over the whole mass range. We consider
enrichment due to PN, SN~II, and SN~Ia (see also eqn.~\ref{eq:all_enrich}).
Since TNH96 models are computed only for solar metallicities,
we consider WW95-yields of solar initial metallicity. The stellar lifetimes are
taken from Schaller \etal\ \shortcite{SSMM92}. For the following
computations we have extrapolated TNH96 yields to $11~\Msun$, and neglected
the contribution from SN~II coming from stars with mass in the range
$8-11~\Msun$ (see WW95). The fraction of close binary systems is $A=0.035$.
As discussed in section~\ref{sec:solar} this value
is calibrated in the chemical evolution model of the solar
neighbourhood. Since the value of this parameter is very small,
the yields of SN~II are only marginally affected by the choice of $A$.
The exact number becomes important
when star formation timescales of $\sim 10$ Gyr and enrichment due to SN~Ia
are considered. For the remnant masses, we adopt Renzini \& Voli (1981) up
to $8~\Msun$, and either WW95 or TNH96 from $11~\Msun$ to $40~\Msun$. In the
range $8-11~\Msun$ the mass of the remnant is taken to be $1.4~\Msun$.

\subsection{Magnesium}
\begin{figure}
\psfig{figure=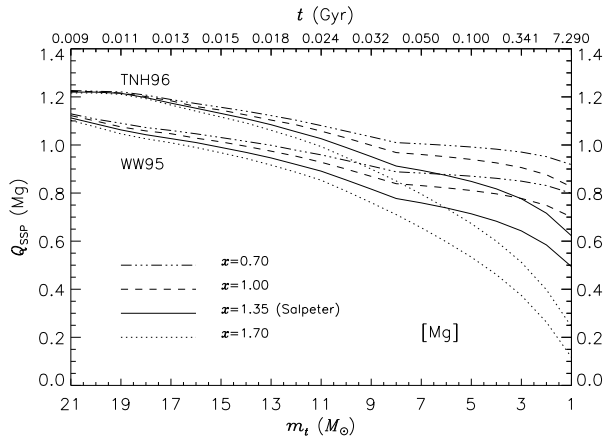,width=\linewidth}
\caption{The figure shows the abundance of magnesium in the ejecta of one
generation of stars in the range from the turnoff to the maximal mass
$m=40~\Msun$. This mass range is increasing with decreasing turnoff mass $m_t$
(lower x-axis) and with increasing time (upper x-axis).
The quantity $Q_{\rm SSP}$ is defined in equation~\ref{time_SSP}.
The
enrichment due to type II SNe ($11-40~\Msun$), type Ia SNe ($3-16~\Msun$) and
planetary nebulae ($1-8~\Msun$) is taken into account. The fraction $A=0.035$ of
binaries exploding as SN~Ia is determined in the chemical evolution model
for the solar neighbourhood in section~\ref{sec:solar}.
The calculated SSP-yield (see eqn.~\ref{time_SSP})
is normalized on the solar magnesium abundance (Anders \& Grevesse 1989)
and plotted on a logarithmic scale. Different SSP-yields are calculated for
different IMF slopes $x$ and SN~II-yields (TNH96, WW95(B,$\Zsun$)). The
total contribution of type II SNe to the SSP-yield is reached after 24 Myr
when the turnoff mass is $11~\Msun$. Since mainly high mass stars contribute
to the enrichment of magnesium, the SSP-yield is decreasing with decreasing
turnoff mass. For Salpeter-IMF, the abundance of 
magnesium in the ejecta of SN~II is $0.13$ dex higher using TNH96-yields.}
\label{sspt_mg}
\end{figure}

Figure~\ref{sspt_mg} shows the abundance of magnesium in the ejecta of one dying
generation of stars as a function of turnoff mass. The abundances are
normalized to solar values and plotted on a logarithmic scale. 
The magnesium abundance in the ejecta is
significantly super-solar. The upper x-axis shows the progression of time
which is not linear with the turnoff mass. The turnoff of $3~\Msun$
is reached after 0.341 Gyr, but it takes more than 7 Gyr until stars of
$1~\Msun$ contribute to the enrichment as well. The different line styles belong
to various IMF slopes. The solid line indicates the Salpeter-IMF.

One can see that the magnesium abundance in the total ejecta is decreasing
with turnoff mass for $m_t<20~\Msun$. 
This is due to the fact that most magnesium is processed
in stars more massive than $20~\Msun$ (figure~\ref{iron_mag}). 
The SN~II-SSP-yield is reached at $m_t=8~\Msun$. For $m_t\leq 8~\Msun$,
SN~Ia and PN begin to contribute. But since both events do not eject a
significant amount of magnesium (see table~\ref{tab:SNIa-yields}), 
the abundance is still decreasing with decreasing turnoff mass and increasing 
time. 

The most striking aspect of this diagram is that the magnesium abundance due
to TNH96 nucleosynthesis is $0.13$ dex higher (Salpeter IMF, model B). 
This is caused by
the significant difference in the yields of $18-25~\Msun$ stars as shown in
figure~\ref{iron_mag}. Hence, the
discrepancy between WW95 and TNH96 is maximum at $m_t\approx 19~\Msun$. 

\subsection{Iron}
\begin{figure}
\psfig{figure=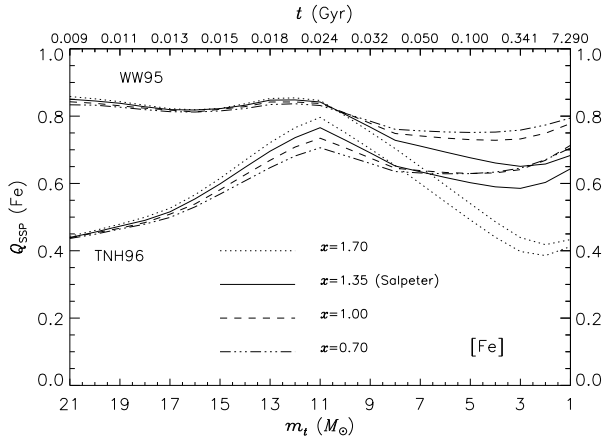,width=\linewidth}
\caption{The figure shows the abundance of iron in the ejecta of one
generation of stars in the range from the turnoff to the maximal mass
$m=40~\Msun$. For a detailed description see the caption of figure~\ref{sspt_mg}.
Since low mass stars dominate the iron yield of SN~II, the maximal
SN~II-SSP-yield is reached at $m_t=11~\Msun$. The value decreases with the
contribution of PNe and rises again when type Ia SNe enter the game. For
Salpeter-IMF, the SN~II-yields of WW95 lead to an iron abundance which is
$0.08$ dex higher.}
\label{sspt_fe}
\end{figure}

The iron enrichment of the SSP as a function of time is shown in
figure~\ref{sspt_fe}.
Since iron is mainly synthesized in stars of
lower masses, the time dependent SSP-yield is roughly constant (WW95) or even
increasing (TNH96) with increasing turnoff mass (fig.~\ref{sspt_fe}) until
the SN~II value is reached. Since stars between
$8-11~\Msun$ are assumed not to contribute to the enrichment of heavy elements, 
there is a peak at $11~\Msun$.
At late times, the contribution due to
type Ia SNe comes into play, and the iron abundance
in the ejecta is rising again. It is important to recognize that the iron
abundance in the ejecta of SN~II is higher for a steeper IMF (see dotted
curve). The results from TNH96-models are more strongly dependent on the slope 
of the IMF than WW95, because the contribution of high mass stars to the
iron production is smaller in TNH96 (see also figure~\ref{cont_mgfe}). 
For the same reason, the difference between
WW95 and TNH96 increases for a flatter IMF. 

\subsection{[Mg/Fe]}
\begin{figure}
\psfig{figure=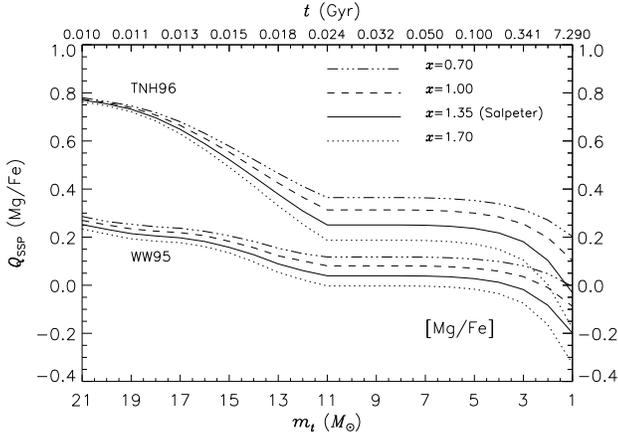,width=\linewidth}
\caption{The figure shows the abundance ratio of magnesium over iron in the
ejecta of one
generation of stars in the range from the turnoff to the maximal mass
$m=40~\Msun$. For a detailed description see the caption of figure~\ref{sspt_mg}.
Since stars between 8 and $11~\Msun$ do neither eject magnesium nor iron,
the ratio is constant in this mass range. The further ejection of iron due to
SN~Ia drives the ratio down for lower turnoff masses. The ratio provided by 
SN~II-yields of WW95 does not suffice to explain magnesium-enhanced
abundance ratios, also for a rather flat IMF.}
\label{sspt_mgfe}
\end{figure}
\begin{figure}
\psfig{figure=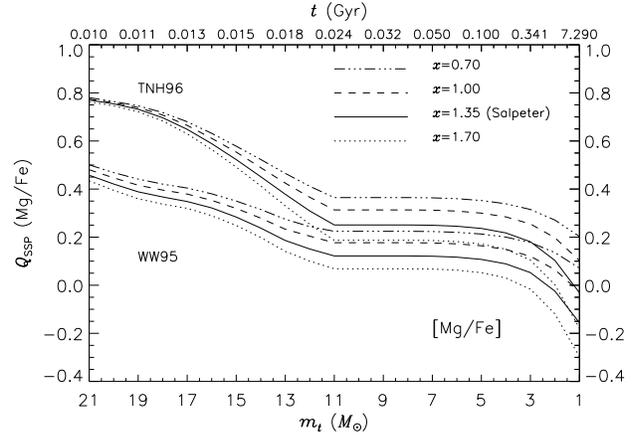,width=\linewidth}
\caption{The diagram shows the same as figure~\ref{sspt_mgfe}. In addition, we
assumed a reduced WW95-iron yield for masses above $20~\Msun$ by a factor of 2, 
according to
the uncertainty given in WW95. The iron yields of masses below $20~\Msun$
are not overestimated by a factor of 2 in WW95, because they agree with
the observations of SN1987A and SN1991J (see text). The SN~II-SSP-yield of iron
is increased by $0.08$ dex, which is not enough to improve the situation
significantly.}
\label{sspt_mgfe2}
\end{figure}

Figure~\ref{sspt_mgfe} shows the following aspects:
\begin{enumerate}
\item In the first 10 Myr, the produced magnesium
overabundance is fairly high, the difference between WW95 and TNH96 is
extremely large. WW95 yields reach [Mg/Fe]$\approx 0.2$
at a turnoff $m_t\approx 20~\Msun$, after 10 Myr. Even considering a
flat IMF with $x=0.70$, the minimum overabundance in ellipticals of 0.2 dex
\cite{WFG92} is reached at $t\approx 15$ Myr when the contribution of
type~II SNe is not yet complete.
TNH96 provide the same value of [Mg/Fe] after 7.3 Gyr when
SN~Ia explosion are already tearing the ratio down. Obviously, this strongly
affects the timescales
of star formation of a system showing a [Mg/Fe] overabundance.
\item One generation of SN~II exploding stars cannot produce the magnesium
overabundance in metal-poor stars in the solar neighbourhood, when
considering Salpeter IMF and WW95-SN~II yields. Assuming a value of
[Mg/Fe]$\approx 0.3$ -- which is already a lower limit -- all low-metallicity
stars should have been born in the first 9 million years (see also
section~\ref{sec:solar}).
\item With progression of time, the overabundance
is decreasing, because more and more low mass stars ($m<20~\Msun$)
with higher iron and lower magnesium yields are contributing to the enrichment. 
Once the turnoff mass is $8~\Msun$, the final SN~II-SSP value is reached.
\item The magnesium overabundance in the SN~II output increases according to
flattening of the IMF. This is because of giving more weight to magnesium
producing high mass stars. 
\item The dependence of [Mg/Fe] on the IMF slope is increasing with time.
This is understandable, because for a larger considered mass range
the role of the IMF slope becomes more important.
\item The TNH96-models provide a [Mg/Fe]-ratio of 0.26 dex, 
WW95-models lead to [Mg/Fe]=0.05 dex, both for Salpeter-IMF.
\end{enumerate}
WW95 specify an uncertainty of a factor of 2 in the iron
yield. Since this causes a shift by 0.3 dex, one could argue that simply
taking half of the iron yield would solve the overabundance problem. We want
to show that this is not the case. Since the iron yields of stars below
$20~\Msun$ are very similar in the two sets of models, and since both
sets reproduce the observations of SN1987A and SN1993J very well (see
table~\ref{tab:ni_obs}), it is reasonable only the halve the iron yields of
stars above $20~\Msun$ in WW95. Re-calculating the SSP-yields with the
modified Fe-yield of WW95, it turns out that the total [Fe] is only shifted
by 0.08 dex to lower values.
Figure~\ref{sspt_mgfe2} shows the [Mg/Fe]-ratio as a function of
the turnoff mass in this experiment. The plot shows, that it remains difficult 
to reproduce the observed magnesium overabundances with WW95 nucleosynthesis.

\smallskip
It is important to mention
that also TNH96 magnesium yields may not suffice to explain observed [Mg/Fe]
overabundances in elliptical galaxies.
There are several indications that [Mg/Fe] in nuclei of ellipticals does
even exceed 0.4 dex \cite{WPM95,Metal97}.
As demonstrated in figure~\ref{sspt_mgfe}, this value can not be theoretically 
produced by one SN~II-exploding
generation of stars. Hence, claiming [Mg/Fe]$\ga 0.4$ dex, the star forming
phase in giant ellipticals must
be of the order $10^{7}$ yr, even for TNH96-yields and a flat IMF ($x=0.7$).
A detailed exploration of star formation timescales, IMF-slopes and stellar 
yields in elliptical galaxies will be the subject of a forthcoming paper.

\smallskip
Finally, one should not forget, that the number of input parameters in the
calculations is very small. 
The above conclusions do not depend on galaxy formation scenarios, 
on star formation histories, on infall models, or on binary fractions.
The only considered parameters are the IMF slope and stellar yields. 

\subsection{[O/Fe]}
\begin{figure}
\psfig{figure=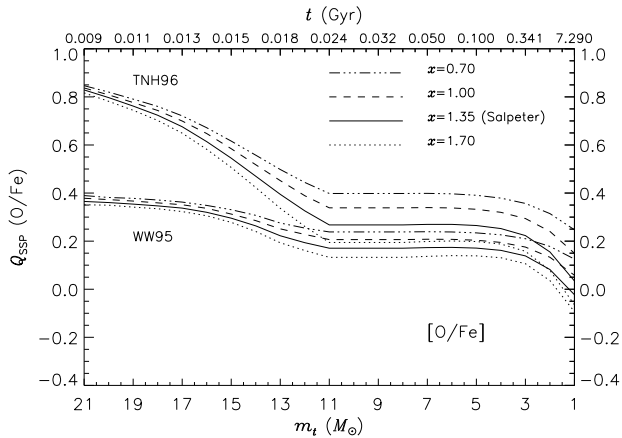,width=\linewidth}
\caption{The figure shows the abundance ratio of oxygen over iron 
in the ejecta of one
generation of stars in the range from the turnoff to the maximal mass
$m=40~\Msun$. For a detailed description see the caption of figure~\ref{sspt_mg}.
Oxygen and magnesium overabundances in the ejecta of SN~II are the same
for TNH96-yields. With WW95, instead, the SSP-value of [O/Fe] is $\sim 0.12$
dex higher than [Mg/Fe]. Since both are $\alpha$-elements and should therefore
be enhanced by approximately the same amount, this is a further indication 
that WW95 underestimate the magnesium yield of SN~II.}
\label{sspt_oxfe}
\end{figure}

Figure~\ref{sspt_oxfe} shows the time dependent SSP yield as a function of
turnoff mass for the abundance ratio [O/Fe]. Since both, oxygen and
magnesium, are produced mainly in type II supernovae, one would expect
similar values for the overabundance.

For TNH96, this is exactly the case. 
The contribution of low mass stars ($1~\Msun\leq M\leq 8~\Msun$)
to the enrichment of oxygen manifests itself in an elongation of the SN~II
plateau down to $5.5~\Msun$. Although SN~Ia produces 6 times more oxygen than
magnesium, the SN~Ia tears the curve down by approximatly the same amount
because of the dominant role of iron in the ejecta.
The WW95-[O/Fe]-ratio, instead, is 0.12 dex higher than [Mg/Fe].

The discrepancy
between WW95 and TNH96 [O/Fe]-ratios now originates mainly from the discrepancy 
in the iron yields, while oxygen yields differ by 0.02 dex. This leads again to 
the conclusion, that WW95 may underestimate the magnesium yield.

\subsection{On the upper mass cutoff}
\label{sec:cutoff}
In this paragraph, we will investigate the influence of a variation of the
upper mass cutoff on the calculated SSP-yields. For this purpose, we include
the results of TNH96 for the $70~\Msun$-star. In order to compare the
different nucleosynthesis prescriptions, we have to extrapolate the
WW95-yields to higher masses, hence the result has to be interpreted with
cautions. In the WW95-models most heavy elements re-implode for the massive 
stars, so that the contribution of these stars to the enrichment is
negligible. Indeed, the plots in figures~\ref{oxy_z} and~\ref{iron_mag} show 
this trend for
the elements oxygen and magnesium, respectively (also model B). TNH96 do not
consider fall back, thus their O-Mg-yields increase with mass up to $m=70~\Msun$.

\begin{figure}
\psfig{figure=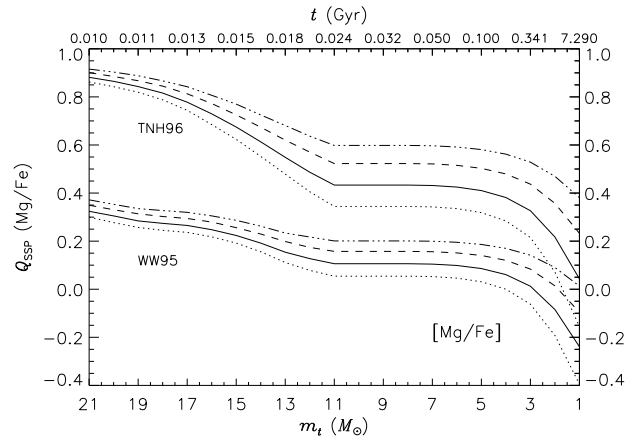,width=\linewidth}
\caption{The figure shows the abundance ratios of magnesium over iron in the
ejecta of one
generation of stars in the range from the turnoff to the maximal mass
$m=70~\Msun$. For a detailed description see the caption of figure~\ref{sspt_mg}.
The consideration of a higher maximum mass leads to larger [Mg/Fe]-ratios
for TNH96-models. The results for WW95 basically do not change. However, the
WW95-yields are extrapolated to $70~\Msun$ and are therefore uncertain.}
\label{fig:sspt_mgfe3}
\end{figure}

Table~\ref{tab:delta} gives the variation of the {\em abundances} of various
elements in the SN~II ejecta of one SSP, if $m_{\rm max}=70~\Msun$ with
respect to the $m_{\rm max}=40~\Msun$ case. These reflect both the metal
production and the total ejected mass in the range $40-70~\Msun$.
\begin{table*}
\begin{minipage}{16cm}
\caption{The numbers give the shift of the abundances in the ejecta of one
SSP when the maximum mass is increased from $40$ to $70~\Msun$. The yields
of WW95 are extrapolated to masses above $40~\Msun$.}
\begin{tabular}{lrrrrrrrrrrrr}
\hline
      & \multicolumn{3}{c}{$x=1.7$} & \multicolumn{3}{c}{$x=1.35$} &
\multicolumn{3}{c}{$x=1.0$} & \multicolumn{3}{c}{$x=0.7$} \\\\
      & $\Delta$[O] & $\Delta$[Mg] & $\Delta$[Fe] & $\Delta$[O] & $\Delta$[Mg] & 
      $\Delta$[Fe] & $\Delta$[O] & $\Delta$[Mg] & $\Delta$[Fe] & $\Delta$[O] & 
      $\Delta$[Mg] & $\Delta$[Fe] \\
TNH96 & $0.20$ & $0.14$ & $-0.01$ & $0.23$ & $0.17$ & $-0.01$ & $0.26$ &
$0.19$ & $-0.01$ & $0.28$ & $0.22$ & $-0.02$ \\
WW95 & $0.09$ & $0.02$ & $-0.03$ & $0.11$ & $0.02$ & $-0.05$ & $0.12$ &
$0.01$ & $-0.06$ & $0.13$ & $0.00$ & $-0.09$ \\
\hline
\end{tabular}
\label{tab:delta}
\end{minipage}
\end{table*}
The following striking aspects should be mentioned:
\begin{enumerate}
\item The iron abundance in the ejecta decreases for all models. This effect
is strongest for flatter IMF and WW95-models.
\item The oxygen abundance, instead, increases for all models. Again the
effect is strongest for a flatter IMF, but more important in TNH96.
\item The behaviour of the magnesium abundance is more complex. For TNH96
the increase of magnesium becomes more significant with a flatter IMF,
whereas for WW95 the pattern is inverse.
\end{enumerate}
In total, the [Mg/Fe]-ratio in the SSP-ejecta increases significantly only
for TNH96-models. In the WW95-models, the effect of fall back prevents a 
significant change.
This is confirmed in figure~\ref{fig:sspt_mgfe3} in which we show  the abundance
ratios as a function of turnoff mass and time with $m_{\rm max}=70~\Msun$
considered.
The diagram shows that assuming a larger value for $m_{\rm max}$ 
the problem of the magnesium overabundance is relaxed, {\em if the
high Mg-yield calculated by TNH96 in high mass stars is correct.} This seems
to us still controversial. It is of great importance in the future to
improve the knowledge on the stellar yields of these stars, too.

Furthermore, it is worth noting that the abundance ratio of magnesium to
oxygen decreases with the inclusion of stars more massive than $40~\Msun$. 
This is
important because it becomes even more difficult to reproduce the solar Mg/O
ratio. We give a detailed exploration on this aspect in
paragraph~\ref{par:cutoff} of the next section.


\section{The solar neighbourhood}
\label{sec:solar}
\subsection{The model parameters}
The chemical evolution of our galaxy is treated in the literature several
times (e.g.\ Matteucci \& Greggio 1986; Timmes \etal~1995; Pagel \&
Tautvaisiene 1995; Tsujimoto \etal\ 1995, Yoshii \etal\ 1996). The model
predictions fit the data quite well, the main observational
features can be reproduced. In the classical numerical models, 
the chemical evolution of the ISM in the solar
neighbourhood is described in a one-zone model of homogeneous and 
instantaneously mixing gas. The latter assumption is called the
instantaneous {\em mixing} approximation. The instantaneous {\em recycling}
approximation, which neglects the stellar lifetimes is relaxed in these
models as well as in our calculations. In principle, the accretion of
gaseous matter over a timescale of $\sim 4$ Gyr enables us to avoid the
formation of extremely metal-poor stars with [Fe/H]$<-3$, known as the G-dwarf
problem \cite{L72,T88,MF89}. 
The formation of the disk in the solar vicinity due
to accretion $f(t)$ is described in the following equation \cite{TWW95}:
\[ f(t)=[M_{\rm tot}(t_{\rm now})-M_{\rm tot}(t_0)] \]
\begin{equation}
\times \frac{\exp(-t/\tau_{\rm disk})}
{\tau_{\rm disk}[1-\exp(-t_{\rm now}/\tau_{\rm disk})]}\ ,
\label{eq:infall}
\end{equation}
with $M_{\rm tot}(t_{\rm now}=15\ {\rm Gyr})$ and $M_{\rm tot}(t_0=0\ {\rm Gyr})$
as the surface densities ($\Msun\spc$)
of the total mass (stars+gas) today and at the beginning of the disk
formation, respectively.
The accretion timescale for the formation of the disk is controlled by the 
parameter $\tau_{\rm disk}$. 

The SFR is assumed to depend on the gas density of the ISM \cite{S59,S63} 
with $\nu$ ($\pGyr$) as the efficiency of star formation (free parameter).
\begin{equation}
\psi(t)=\nu\: M_{\rm tot}\left[\frac{M_{g}(t)}{M_{\rm tot}(t)}\right]^k
\label{eq:SFR}
\end{equation}
In the literature, the adopted value for the exponent $k$ varies between
$k=1$ and $k=2$ (e.g.~Matteucci \& Fran\c{c}ois 1989). 
In the following section, we will show the influence of this 
parameter on the observational features.
The Schmidt-law together with the infall of gas over a relatively long 
timescale guarantee a roughly continuous star formation during the evolution 
of the solar neighbourhood.

Furthermore, the enrichment of the ISM due to PN, SN~II and SN~Ia is
considered, using supernova rates as described in section~\ref{sec:general}.
The parameter $A$ in equations~\ref{eq:SNII} and ~\ref{eq:SNIa}
is a free parameter. It is calibrated on the current supernova rates in our
Galaxy.
As shown in the previous sections, especially the yields of type II SNe are
affected by many uncertainties. Hence, we treat the SN~II-yields as a
parameter in the sense that we consider the different SN~II nucleosynthesis
prescriptions presented in section~\ref{sec:yields} (WW95 and TNH96).
TNH96-yields consist of the given yield from the evolution of the helium
core plus the initial abundance of the element in the envelope (see also
section~\ref{sec:yields}). In the simulations of the chemical evolution, 
the initial element abundances of the envelopes correspond to the element 
abundances in the ISM when the star forms. In these terms, the TNH96-yields
become metallicity dependent, although the evolution of the helium 
core is only calculated for solar element abundances.

The basic equations of chemical evolution are explained in 
section~\ref{sec:general}. Since SN~Ia explode delayed with respect to SN~II
\cite{GR83}, 
the element abundances in metal-poor stars are
determined mainly by SN~II. Hence, the adopted standard model for the chemical
evolution in the solar neighbourhood can easily explain the enhancement of 
$\alpha$-elements in metal-poor stars, assuming the [Mg/Fe]-ratios given by
SN~II nucleosynthesis are high enough.

\subsection{Observational constraints}
There are several observational features in the solar neighbourhood basically 
constraining different parameters. In the subsections below, we will discuss
in detail the influence of the parameters on the abundance distribution
function (ADF), the age-metallicity relation (AMR), the current supernova
rates, and the element abundances in the sun. The parameters have to be
adjusted to provide the best possible {\em simultaneous} fit to the existing 
observational data. In table~\ref{tab:parameters} we summarize how the various
parameters can be constrained by the different observational features.
The right column of the table gives the final adopted values.
\begin{table*}
\begin{minipage}{12cm}
\caption{Input parameters in the calculations for the chemical evolution of
the solar neighbourhood. The parameters are chosen to match simultaneously
the observational constraints: ADF, AMR, supernova rates, solar
element abundances, current infall rate, and current fraction of gaseous mass. 
The second column shows the main observational constraints on the respective
parameter. The third column gives the final adopted values.}
\begin{tabular}{lll}
\hline
Parameter & Observational constraint & Adopted value \\\\
Stellar yields & Element abundances of the sun & TNH96 \\
IMF slope $x$ & Solar abundance ratios & 1.36 \\
Close binary fraction $A$ & Relative frequency of type II and Ia SNe  & 0.035 \\
                          & AMR & \\
Star formation efficiency $\nu$ & current fraction of gaseous mass  & $1.3~\pGyr$\\
Schmidt exponent $k$ & ADF  & 2 \\
Accretion timescale $\tau_{\rm disk}$ & current infall rate & 4 Gyr \\
                                      & ADF & \\
\hline
\end{tabular}
\label{tab:parameters}
\end{minipage}
\end{table*}
The calculations are performed using the stellar yields of TNH96.
Additional computations for WW95-yields under the same conditions are made
in order to work out the influence of stellar nucleosynthesis. 
The galactic age is assumed to be $t_{\rm now}=15$ Gyr \cite{TWW95}, 
the age of the
sun is 4.5 Gyr. The value of the surface density in the solar 
neighbourhood is assumed to be $77 ~\Msun\spc$
\cite{KG89a,KG89b,KG89c,S89,G90,KG91}.
Stellar lifetimes are taken from Schaller \etal\ \shortcite{SSMM92}.

\subsubsection{Abundance Distribution Function}
\label{sec:adf}
The differential ADF gives the
number of stars that are born per unit metallicity as a function of
metallicity. Pagel \& Patchett \shortcite{PP75} derived this relation for
the solar vicinity ($\sim 25$ pc) with a sample of 132 G-dwarfs.
The most important feature of the ADF
is the paucity of extremely metal-poor stars. Assuming a closed box for the
chemical evolution, the so called {\em Simple Model} predicts too many 
metal-poor stars \cite{vdB62,S63,T80}. This deviation is known as the G-dwarf
problem. The considerations of pre-enrichment \cite{TC71} or infall of material 
\cite{L72} help to
avoid the formation of low-metallicity stars. The latter possibility is used
in the adopted model for the solar neighbourhood, assuming the disk to form
due to accretion of primordial gas (see equation~\ref{eq:infall}).

\begin{figure}
\psfig{figure=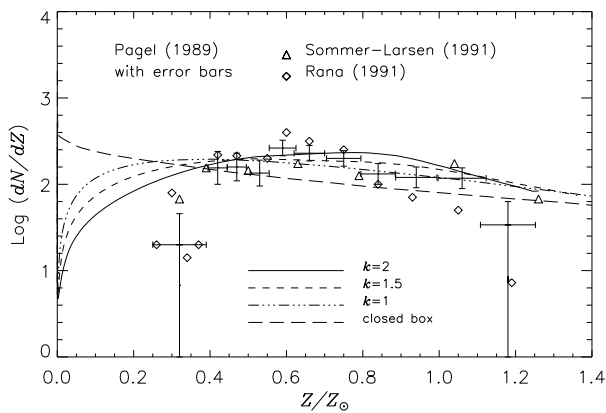,width=\linewidth}
\caption{The abundance distribution function (ADF) giving the number of stars
that are born per unit metallicity $\log (dN/dZ)$ as a function of metallicity.
The observational data points with error bars refer to the reanalysis of the
Pagel \& Patchett data (1975) by Pagel (1989), taking the metallicity-excess
calibration of Cameron (1985) into account. Additional re-interpretations of
the data set by Rana (1991) and Sommer-Larson (1991) respective
with correction for the increase of the velocity dispersion with time and
the correction for the vertical height distribution of dwarfs. The long-dashed
line shows the calculated ADF for a closed box model without infall. While
this model definitly fails to match the observations, the models with the
inclusion of infall can reproduce at least the general shape of the observed ADF.
The best fit refers to the exponent $k=2$ of the Schmidt-law for fixed
accretion timescale $\tau_{\rm disk}=4$ Gyr. The parameters $\nu$ and $k$ 
are chosen such that the same amount of gas is converted to stellar mass in
all models.}
\label{fig:adf}
\nocite{Pa89,R91,SL91,C85}
\end{figure}

The shape of the resulting theoretical ADF depends basically on dynamical
parameters like the accretion timescale $\tau_{\rm disk}$ and the Schmidt-law
exponent $k$ \cite{MF89}. Figure~\ref{fig:adf} shows the results for
different choices of the parameter $k$.
To garantuee that in all computations the same total number of stars is
formed, the star formation efficiency $\nu$ (see equation~\ref{eq:SFR}) is
reduced for smaller $k$. The diagram demonstrates the following:
\begin{enumerate}
\item The inclusion of infall solves the G-dwarf problem in the sense that
the extremely high amount of metal-poor stars as predicted by the cosed box
model (long-dashed line) is significantly decreased. The general shape of
the ADF, the peak at intermediate metallicities, is reproduced by the model.
\item The smaller the exponent $k$ is, the more stars of
high and low metallicity are formed. Since the ADF predicted by the model
is already too flat, $k=2$ may be the best choice.
\end{enumerate}
A better fit to the ADF-data requires an improvement of the adopted model.
Since there are both, too many metal-poor and metal-rich stars, a different
description of the infall-term may be necessary. In addition, the
consideration of pre-enrichment of the infalling gas further reduces the
number of low metallicity stars. Since the aim of this work is to
inspect the influence of different stellar yields on the chemical evolution
in the solar vicinity in the framework of the standard infall-model model, 
we simply use the ADF to constrain the model parameters without improving
the model to obtain better fits.

\begin{figure}
\psfig{figure=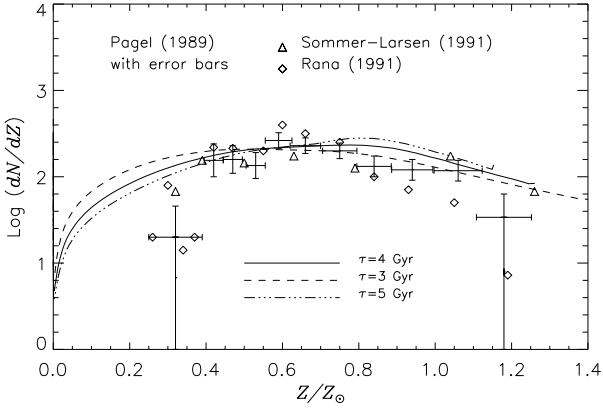,width=\linewidth}
\caption{The abundance distribution function (ADF) giving the number of stars
that are born per unit metallicity $\log (dN/dZ)$ as a function of metallicity.
For a short description of the data points see figure~\ref{fig:adf}. In this
plot, the accretion timescale is varied. Since a longer accretion timescale
supports the formation of metal-rich stars, the best fit to the data is
obtained for $\tau_{\rm disk}=4$ Gyr. The Schmidt exponent is $k=2$ as
worked out above.}
\label{fig:adf_tau}
\end{figure}

Figure~\ref{fig:adf_tau} shows the ADF for different accretion timescales.
For $\tau_{\rm disk}=3$ Gyr and $\tau_{\rm disk}=5$ Gyr the number of 
metal-poor and metal-rich stars is overestimated, respectively. Thus, we use 
$\tau_{\rm disk}=4$ Gyr in our simulations.

\subsubsection{Age Metallicity Relation}
\begin{figure}
\psfig{figure=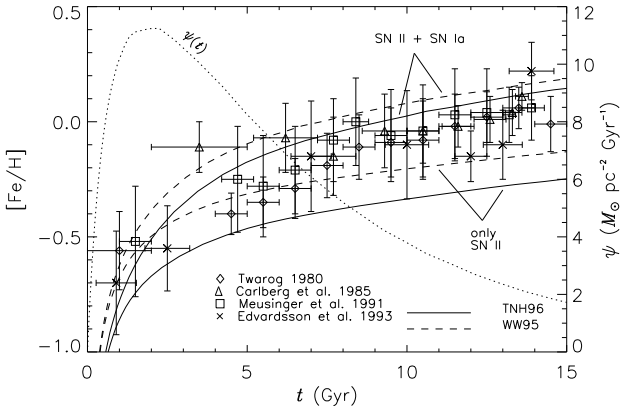,width=\linewidth}
\caption{The age-metallicity relation (AMR) for the solar neighbourhood. The
symbols indicate the observational data, where the error bars denote the
spread of many stars in the data. Twarog (1980) determined age and
metallicity for two samples of 1007 and 2742 local disk stars, respectively.
Carlberg \etal\ (1985) and Meusinger, Reimann, \& Stecklum (1991) reanalysed 
these data using new
isochrones from VandenBerg (1985). Edvardsson \etal\ (1993) did not re-examine 
the Twarog-data but derived abundances for 189 F and G disk dwarfs in the solar
vicinity. The plot shows that for both SN~II nucleosynthesis prescriptions,
the enrichment due to type Ia SNe is necessary to reproduce the AMR in the solar
neighbourhood. A comparison between the solid and the dashed lines (TNH96
and WW95 (model B) SN~II-yields, respectively) confirms the result from
section~\ref{sec:ssp} that WW95 produce $\sim 0.08$ dex more iron. The dotted
curve shows the SFR ($\Msun\spc\pGyr$) as a function of time. The value for
today ($t=15$ Gyr) is in agreement with observations (G\"usten \& Mezger
1983).}
\label{fig:amr}
\nocite{VdB85,GM83,MRS91,CDHV85}
\end{figure}

The age metallicity relation (AMR) shows the ratio [Fe/H] indicating the
metallicity as a function of the ages of the stars \cite{Tw80}. Since different
element abundances of the ISM at different times are locked in stars of
different ages, this corresponds to the evolution of [Fe/H] in the ISM as a
function of time. In the first 2 Gyr of the evolution when the SFR is at its
maximum, [Fe/H] rises very steeply to a value of $\sim -0.5$ dex. The
increase flattens out significantly and emerges to solar metallicity at
$t\approx 10$ Gyr. Figure~\ref{fig:amr} shows, that this behaviour is well
reproduced by the simulations. Star formation (dotted line)
is occurring over the
whole range of 15 Gyr with a peak of $11~\Msun~\spc~\pGyr$ at $t=1.9$ Gyr.
Fitting an exponential law like
\[ \psi(t)\sim\exp^{-t/\tau}\]
to the range $5-15$ Gyr where the SFR is decreasing leads to $\tau\approx 8$~Gyr.

The iron abundance at large $t$ is significantly determined by the
contribution of type Ia SNe. The simulations excluding enrichment by SN~Ia
clearly underestimate the production of iron.\footnote{This statement is
basically independent of $m_{\rm min}$, since the abundance {\em ratio} of
iron to oxygen is underestimated without SN~Ia.} Furthermore, the fraction of
iron contributed by the different types of SNe depends on the adopted
stellar yields.
WW95 models (dashed lines) produce $\sim 0.08$ dex more iron than TNH96 
(see section~\ref{sec:ssp}). 
Using TNH96 yields, 60 per cent of the produced iron comes from type Ia 
SNe, according to WW95 models this amount decreases to 50 per cent. 
Because of the higher iron yield, WW95 models fit the AMR-relation worse,
but are still within the error bars.

The total amount of iron in the ISM highly depends on the
fraction $A$ of close binary systems. The AMR could
be slightly better fitted for a reduced iron production, thus for a lower
parameter $A$. However, this parameter is additionally constrained by the
relative frequency of the different types of supernovae.

\subsubsection{Supernova rates}
Unfortunately, the current rates of both types of SNe (Ia and II) in spiral 
galaxies and in the solar neighbourhood are still uncertain \cite{vBT91}.
Since there is no consensus, the range allowed by observations is
fairly large. 
The theoretical relative frequency of SN~II and SN~Ia is mainly determined by the
parameter $A$. In their review paper, van den Bergh and Tammann \shortcite{vBT91}
claim $N_{\rm SNII}/N_{\rm SNIa}\approx 2.7$ for Sab--Sb galaxies and
$N_{\rm SNII}/N_{\rm SNIa}\approx 8$ for Sc--Scd galaxies. Since our Galaxy
is assumed to have a Hubble type between Sb and Sc \cite{vBT91}, a relative
frequency of $N_{\rm SNII}/N_{\rm SNIa}\approx 5$
seems to be a reasonable estimate. This value is in agreement with our
calculations. 

\begin{figure}
\psfig{figure=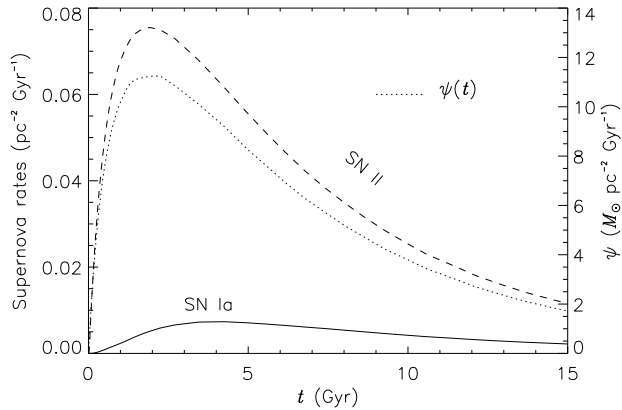,width=\linewidth}
\caption{The plot shows the rates of both types of supernovae and the SFR as
a function of time. The enrichment due to SN~Ia is delayed with respect to
SN~II. The dotted curve demonstrates that the rate of type II SNe -- occuring
in high mass and short-living stars -- directly depends on the SFR. The
calculated number ($\spc\pGyr$) of SN~II occuring today is in rough agreement 
with observational estimates (see text). The
relative frequency of SN~II to SN~Ia is highly dependent on the parameter
$A$ and is in agreement with observations (van den Bergh \& Tammann
1991). The fraction $A$ of close binaries has to be chosen to fit the 
supernova rates and the AMR (figure~\ref{fig:amr}) simultaneously.}
\label{fig:sn-rates}
\end{figure}

Figure~\ref{fig:sn-rates} shows the rates of type II and type Ia SNe as a
function of time.
While the relative frequency of SN~II and SN~Ia basically constrains the
parameter $A$, the absolute number of type II SNe occuring today depends on
the parameters $\nu$, $k$, and $\tau_{\rm disk}$, whose values are already
chosen to fit the ADF. Assuming that SNe~II occur in
stars above $8~\Msun$, Tammann \shortcite{T82,vBT91} estimates a surface
density of $N_{\rm SNII}\approx 0.02~\spc\pGyr$ in the solar neighbourhood
from historical data. However, because of the small size of the sample, 
this value is quite uncertain. Indeed,
van den Bergh \shortcite{vBT91} 
claims that the historical data may overestimate the absolute
number of SN~II significantly. Thus, the calculated value of
$N_{\rm SNII}\approx 0.01~\spc\pGyr$ is still acceptable.

\subsubsection{Solar element abundances}
The model assumes that disk formation started 15 Gyr ago. Since the sun is
$\sim 4.5$ Gyr old, the element abundances in the ISM predicted by the model
have to be solar at $t\approx 10.5$ Gyr. 

In section~\ref{sec:general}, we showed how to constrain the SFR (parameters
$\nu$ and $k$), the IMF slope ($x$) and the stellar yield from observational
data of the accretion rate, the current gas fraction and the element abundances.
The timescale for disk
formation $\tau_{\rm disk}$ is constrained by the accretion rate which is
observed today (see caption of table~\ref{tab:constraints})\footnote{The
today's accretion rate of $0.2-1.0~\Msun\spc\pGyr$ estimated from
observations of high velocity H$\,${\sc i} clouds (see Timmes~\etal\ 1995
and references therein) allows $\tau_{\rm disk}\approx 3-5.5$ Gyr. The more
specific value of 4 Gyr is constrained by the ADF.}.
Having fixed $f(t)$ and $m_{\rm min}$, the current fraction of gaseous mass 
constrains the mean SFR ($\rightarrow\nu,k$), depending on the IMF slope.
We showed that, considering an
element whose yield is relatively certain, the calculated solar abundance of
this element
depends on $\bar{\psi}$ and $x$. Thus, $\tau_{\rm disk}$, $\nu$, $k$, and $x$ are
fixed. 

Figure~\ref{oxy_z} shows that WW95 and TNH96 differ only slightly in the
calculated oxygen yield. Furthermore, oxygen is mainly produced in massive
stars of small stellar lifetimes $\tau_m$, 
thus the neglection of $\tau_m$ in the 
arguments in section~\ref{sec:general} is valid.
Hence, we assume this yield to be the most certain
and use oxygen to pin down the IMF slope. Having done this, we
can analyse if the stellar yields of various elements are in agreement with
observations.
Table~\ref{tab:constraints} shows the comparison between the 
calculated quantities and their observational constraints.
\begin{table}
\caption{Numerical results of the chemical evolution in the solar
neighbourhood compared with observational constraints. The adopted input
parameters are given in table~\ref{tab:parameters}.
The current fraction of gas $M_g/M_{\rm tot}(t_{\rm now})$ is taken from 
Rana \& Basu (1992), the current accretion rate ($\Msun\spc\pGyr$) comes from 
observations of high velocity H$\,${\sc i} clouds (see Timmes~\etal\ 1995 
and references therein). Solar element abundances (by mass) are adopted 
from Anders \& Grevesse (1989), meteoritic values.}
\begin{tabular}{llll}
\hline
                      & TNH96       & WW95        & Observation    \\\\
$M_g/M_{\rm tot}(t_{\rm now})$ 
                      & 0.13        & 0.13        & $0.10\pm 0.03$   \\
$f(t_{\rm now})$      & 0.46        & 0.46        & $0.2-1.0$      \\\\
Solar $Z$             & $1.96(-2)$  & $1.86(-2)$  & $1.88(-2)$  \\
Solar $\nucl{1}{H}$   & $6.96(-1)$  & $6.89(-1)$  & $7.06(-1)$  \\
Solar $\nucl{16}{O}$  & $9.92(-3)$  & $9.36(-3)$  & $9.59(-3)$  \\
Solar $\nucl{24}{Mg}$ & $4.80(-4)$  & $3.68(-4)$  & $5.15(-4)$  \\
Solar $\nucl{56}{Fe}$ & $1.26(-3)$  & $1.41(-3)$  & $1.17(-3)$  \\
\hline
\end{tabular}
\label{tab:constraints}
\nocite{AG89,RB92}
\end{table}
The element abundances of $\nucl{1}{H},\;\nucl{16}{O}$ and $Z$ are best
reproduced (due to the above strategy): WW95
and TNH96 differ only marginally, the deviations from observational data are
between 1 and 4 per cent.
However, in the case of magnesium, the situation is different: 
the calculated $\nucl{24}{Mg}$-abundance deviates from observational data by
7 per cent (TNH96) and 29 per cent (WW95). 
Reproducing the solar oxygen abundance, the calculated
magnesium abundance is too low, especially with WW95 nucleosynthesis. 
Hence, the predicted ratio between the two
element abundances is not in agreement with observations.
Since SN~II is the main contributor to the Mg-enrichment, we can conclude
that the magnesium yield of type II SNe is clearly underestimated by WW95.

Although $\nucl{56}{Fe}$ also deviates from the observational value, 
one cannot directly draw conclusions on the iron yield of type II supernovae, 
because of the large contribution due to type Ia SNe.

\subsubsection{[Mg/Fe]}
\begin{figure}
\psfig{figure=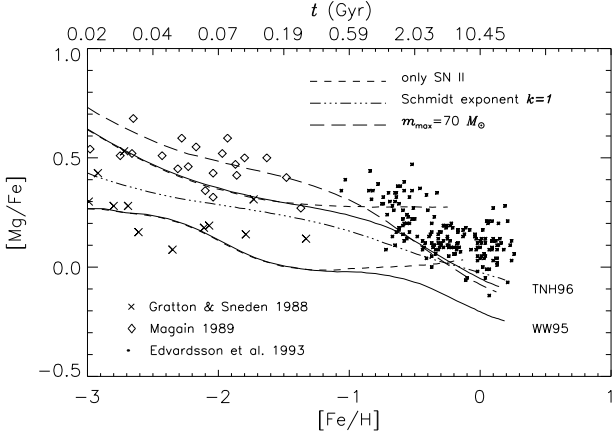,width=\linewidth}
\caption{The abundance ratio [Mg/Fe] as a function of [Fe/H] and time (upper
x-axis). Magain (1989)
derived element abundances in 20 metal-poor halo stars leading to a mean
[Mg/Fe] of $0.48$ dex. Gratton \& Sneden (1988) measured the
abundances of 12 metal-poor field giants and derived a mean
[Mg/Fe]$\approx 0.27$ dex. The two solid lines show the results of the
simulations for SN~II-yields of TNH96 and WW95 (model B), respectively,
taking enrichment due to both types of SNe into account. The dashed curve
corresponds to calculations only considering SN~II. The abundance ratio
reaches the SN~II-SSP-value,
once a complete SN~II-generation of stars enriches the ISM
($m_t\leq 11~\Msun$). The value of the `plateau' is approximately
the SSP-value of
SN~II as given in table~\ref{tab:B1.35}. At higher metallicities, the
iron-dominated ejecta of SN~Ia drive the ratio down. The plot shows that the
[Mg/Fe]-ratio in the ejecta of WW95-SNe is too low to explain the
observational data. The dashed-dotted and the long-dashed lines show the results
for $k=1$ and $m_{\rm max}=70~\Msun$, respectively (both TNH96 yields).}
\label{fig:mgfe_fe}
\end{figure}

We now turn to consider the element
abundances observed in stars of various metallicities in the solar
neighbourhood,
being the last important observational constraint on theoretical models.
Gratton \& Sneden \shortcite{GS88} and Magain \shortcite{M89} 
determined [Mg/Fe] in metal-poor
halo stars, Edvardsson \etal\ \shortcite{EAGLNT93} determined [Mg/Fe] in 
disk stars with 
[Fe/H]$\geq -1$. These data together with the theoretical predictions from
the model using the
parameters of table~\ref{tab:parameters} are plotted in figure~\ref{fig:mgfe_fe}.

The scatter of the data for [Fe/H]$<-1$ is extremely large.  A reasonable
average in the range $-3\leq$[Fe/H]$\leq -1$ seems to be
$0.3\leq$[Mg/Fe]$\leq 0.4$ dex. While TNH96-[Mg/Fe]-yields are high enough to
fit this value, WW95 fail to reproduce such large values.
The same preliminary conclusion was already made in section~\ref{sec:ssp}.

\smallskip
Timmes \etal\ \shortcite{TWW95} also realized, that the produced [Mg/Fe]-ratio 
in WW95 is too low to explain the data, and suggested a reduction of the SN~II
iron yield by a factor of 2. On the other hand, as discussed in
section~\ref{sec:ssp}, the iron yields of stellar masses smaller than
$20~\Msun$ are in good agreement with the observational data from SN1987A and
SN1991J. Thus, it is reasonable
to halve the iron yield of stellar masses {\em greater} than $20~\Msun$.
We showed that this results in a shift of 0.08 dex to higher [Mg/Fe]-values,
which is not enough to explain a [Mg/Fe] overabundance of $0.3-0.4$ dex.

Since a flatter IMF would result in an overestimation of the solar
metallicity and oxygen abundance, the observed trend can only be produced with an
increased Mg-yield. Since it is the Mg/O ratio which is underestimated,
a variation of $m_{\rm min}$ is not suitable to solve the problem, either.

Timmes \etal\ \shortcite{TWW95} claim that a small contribution of type Ia 
SNe or intermediate- and low-mass stars to the magnesium enrichment 
could solve the problem without increasing the SN~II-yields. But both
alternatives seem to be unlikely for the following reasons:
\begin{enumerate}
\item Low-mass stars ($1-8~\Msun$) form CO-WD and therefore do not burn
carbon to magnesium \cite{RV81}.
\item Intermediate-mass stars ($8-11~\Msun$) may even produce a lower
[Mg/Fe]-ratio than high-mass stars, because the ratio decreases with
decreasing mass (see figure~\ref{cont_mgfe}). If this
trend can be approximately extrapolated to lower masses, those stars do not
increase the value of [Mg/Fe] in the ISM.
\item SN~Ia may be a candidate for a higher magnesium production. On the
other hand,
the [Mg/Fe]-ratio is underestimated in a regime at low metallicities, 
where type II SN products dominate and type Ia SNe do not play any role.
\end{enumerate}
However, uncertainties in convection theory and stellar evolution are high
enough to cause different Mg-yields of SN~II, which is convincingly 
demonstrated in the discrepancy between WW95 and TNH96.

\smallskip
The observational data points in figure~\ref{fig:mgfe_fe} 
show two different slopes at different metallicity ranges. 
The progression of [Mg/Fe] is very flat in the low [Fe/H]-region and only 
slightly decreasing with increasing metallicity. 
This belongs to the regime in the first 70 Myr (see scale at the upper x-axis), 
where type II supernovae are dominating the enrichment of the ISM. For
[Fe/H]$\ga -1$, type Ia supernovae enter the game and drive down the
[Mg/Fe]-ratio because of their iron dominated ejecta. The decrease of
[Mg/Fe] with increasing [Fe/H] becomes notedly steeper.
Although the theoretical curves reflect this behaviour roughly, using TNH96
yields the slope at low metallicities is still too steep, especially for 
$-3\leq$[Fe/H]$\leq -2$.

This decrease becomes flatter for the smaller Schmidt exponent $k=1$ as
indicated by the dashed-dotted line.
However, while the choice of $k=1$ may improve the agreement with the data in the
[Mg/Fe]-[Fe/H]-diagram, the ADF is worse reproduced (see figure~\ref{fig:adf}).
The offset to the data of $\sim 0.1$ dex seems to have its origin in a too 
low magnesium yield. Since the model parameters are chosen to reproduce the
solar oxygen abundance, it is interesting to consider the [O/Fe] ratio as a 
function of [Fe/H], too.
\subsubsection{[O/Fe]}
\begin{figure}
\psfig{figure=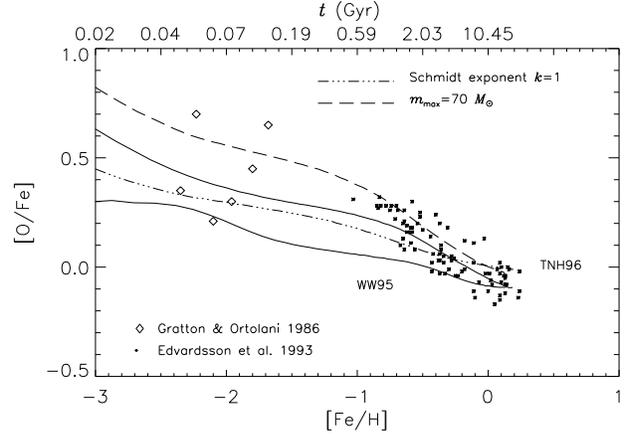,width=\linewidth}
\caption{The abundance ratio [O/Fe] as a function of [Fe/H] and time (upper
x-axis). The two solid lines show the results of the
simulations for SN~II-yields of TNH96 (upper line) and WW95, respectively.
The diagram shows that the [O/Fe]-ratio in the solar neighbourhood is well
fitted by the model. The dashed-dotted and the long-dashed lines show the results
for $k=1$ and $m_{\rm max}=70~\Msun$, respectively (both TNH96 yields).}
\nocite{GO86}
\label{fig:oxfe_fe}
\end{figure}

Again we compare the theoretical results with observations by Edvardsson
\etal\ (1993) at high metallicities and Gratton \& Ortolani
(1986)\nocite{GO86} in the low metallicity regime (figure~\ref{fig:oxfe_fe}). 
These data points are very few and show a large scatter. Thus we concentrate 
on the discussion of the Edvardsson \etal\ data.
These are well fitted by the model using TNH96 nucleosynthesis, hence
the oxygen abundance in the solar neighbourhood is reproduced as well. The
calculation with a lower Schmidt-exponent $k$ (TNH96 yields, dashed-dotted
line) clearly fails to
match the observed [O/Fe]. Although WW95-models suffice to produce a
solar [O/Fe] ratio, they give a bad fit to the data for ${\rm [Fe/H]}<0$.

\subsection{On the upper mass cutoff}
\label{par:cutoff}
As already mentioned, we additionally performed calculations with 
$m_{\rm max}=70~\Msun$ and TNH96 nucleosynthesis. The fitting parameters had
to be re-adjusted, the new values and the results for the calculated solar
abundances are given in table~\ref{tab:const_cutoff}. Again, the parameters
were chosen to match the observational constraints simultaneously.
\begin{table}
\caption{Same as table~\ref{tab:constraints}, considering
different upper mass cutoffs for TNH96 nucleosynthesis. In order to maintain
the agreement with the discussed observational constraints, the following input
parameters had to be re-adjusted: $x=1.5$, $A=0.06$, $\nu=1.1~\pGyr$.}
\begin{tabular}{llll}
\hline
                      & $40~\Msun$  & $70~\Msun$  & Observation    \\\\
$M_g/M_{\rm tot}(t_{\rm now})$ 
                      & 0.13        & 0.13        & $0.10\pm 0.03$   \\
$f(t_{\rm now})$      & 0.46        & 0.46        & $0.2-1.0$      \\\\
Solar $Z$             & $1.96(-2)$  & $1.81(-2)$  & $1.88(-2)$  \\
Solar $\nucl{1}{H}$   & $6.96(-1)$  & $7.17(-1)$  & $7.06(-1)$  \\
Solar $\nucl{16}{O}$  & $9.92(-3)$  & $9.93(-3)$  & $9.59(-3)$  \\
Solar $\nucl{24}{Mg}$ & $4.80(-4)$  & $4.43(-4)$  & $5.15(-4)$  \\
Solar $\nucl{56}{Fe}$ & $1.26(-3)$  & $1.21(-3)$  & $1.17(-3)$  \\
\hline
\end{tabular}
\label{tab:const_cutoff}
\end{table}
\begin{quotation}
{\em The solar Mg abundance is even worse reproduced, since the ratio of 
magnesium to oxygen decreases with the inclusion of $70~\Msun$-stars (see also
table~\ref{tab:delta}).}
\end{quotation}
A further remarkable effect is the stronger influence of type Ia SNe on the
enrichment of iron (parameter $A$), because iron is the only element
(of the considered ones) that is not additionally ejected by extremely high
mass stars. The ratio $N_{\rm SN~II}/N_{\rm SN~Ia}\approx 3$ is still within
the range allowed by observation, it may be even a better fit to the
historical data (see discussion above).

\smallskip
The effect on the [Mg/Fe]-[Fe/H]-diagram is shown by the long-dashed curve
in figure~\ref{fig:mgfe_fe}. The ratio [Mg/Fe] is higher by $\sim 0.1$ dex
in the low metallicity regime, whereas it is still too small at higher
metallicities ($-0.7\leq {\rm [Fe/H]}\leq 0.3$). The long-dashed  line in
figure~\ref{fig:oxfe_fe} shows that the oxygen abundance in the solar
neighbourhood can still be reproduced.

The other observational
constraints are also matched. The differential ADF changes as if there was
a kind of pre-enrichment: it increases more rapidly at the lower $Z$.
However already at $Z/\Zsun=0.05$ it presents too many objects, with respect
to the observations.

\subsection{Delayed mixing}
The upper x-axis in figure~\ref{fig:mgfe_fe} shows that the steep decrease
of the TNH96-curve at low metallicities comes from the short timescales in this
regime. On the other hand, this is a consequence from the IMA, 
assuming that the stellar ejecta mix immediately with
the ISM. Although there is no doubt that this assumption is not realistic
\cite{S63,T75},
most chemical evolution calculations hold this approximation. The validity
of the IMA depends on the timescale of
the mixing process. Malinie \etal\ (1993)\nocite{Metal93} claim that
due to chemical inhomogeneities in the disk, re-mixing and star formation may
be delayed by $10^{8-9}$ yr. We will include the consideration of delayed
mixing in our calculations and inspect the influence of different mixing
timescales on the observational constraints discussed above.

\subsubsection{The two gas phases}
We distinguish between two different phases of the gas component:
the {\em active} and the {\em inactive} phase. The {\em inactive} gas
consists of the enriched stellar ejecta. Since this component is hot and not
homogeneously distributed, stars cannot form out of this phase. The {\em
active} phase, instead, is assumed to be cool and well mixed, hence, star
formation is possible only in the {\em active} gas phase. 
In order to keep the circle of star formation and
chemical enrichment alive, the {\em inactive} phase converts to the {\em
active} star forming phase on a certain timescale, which includes
both the {\em cooling} and the {\em mixing} process. The timescale is
treated as a free parameter in the simulations. 

To include this scenario in the calculations, we modify the
equations~\ref{eq:gas} and~\ref{eq:abundance} presented in 
section~\ref{sec:general}:
\begin{eqnarray}
dM_g^{\rm inactive}/dt &=& E -  \frac{1}{\tau_{\rm mix}}\,M_g^{\rm
inactive} \label{eq:inactive} \\
dM_g^{\rm active}/dt   &=& -\psi + f + \frac{1}{\tau_{\rm mix}}\,
M_g^{\rm inactive} \label{eq:active}
\end{eqnarray}
To keep the equations as simple as possible, 
we assume the mass flow between the two gas phases to be proportional to
the total amount of {\em inactive} gas divided by the mixing timescale.
We now have to distinguish between the abundance in the {\em active} and the
abundance in the {\em inactive} gas phase.
\begin{equation}
M_g^{\rm inactive}\,dX_i^{\rm inactive}/dt =
E_{i}-X_i^{\rm inactive}\,E
\label{eq:abundance_inactive}
\end{equation}
\[ M_g^{\rm active}\,dX_i^{\rm active}/dt = \]
\begin{equation}
(X_i^{\rm inactive}-X_i^{\rm active})\frac{1}{\tau_{\rm mix}}\,M_g^{\rm inactive}
+ (X_{i,f}-X_i^{\rm active})\:f
\label{eq:abundance_active}
\end{equation}
The SFR described by the Schmidt-law is then dependent on the density of the
{\em active} gas:
\begin{equation}
\psi =\nu\: M_{\rm tot}
\left[\frac{M_{g}^{\rm active}}{M_{\rm tot}}\right]^k
\label{eq:SFR_mix}
\end{equation}
The infalling material is assumed to mix instantaneously with the {\em
active} gas.

\subsubsection{Observational constraints}
\begin{figure}
\psfig{figure=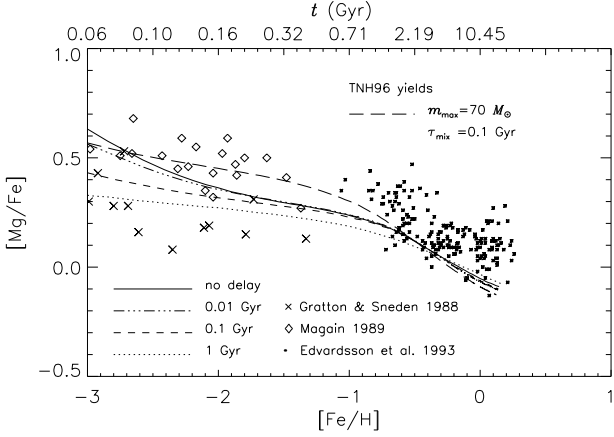,width=\linewidth}
\caption{The abundance ratio [Mg/Fe] as a function of [Fe/H]. For a detailed
description see caption of figure~\ref{fig:mgfe_fe}. In these calculations,
we relax the assumption of the IMA and
consider different timescales for the mixing of the stellar ejecta with the
ISM. SN~II-yields of TNH96 are used. The upper x-axis gives the progression
of time for the case $\tau_{\rm mix}=0.1$ Gyr.
The more the mixing is delayed, the
flatter the curve becomes. The approximately constant value (there might be
a slight decrease) of [Mg/Fe] in metal-poor stars between
$-3\leq {\rm [Mg/Fe]}\leq -1$ can be better reproduced when a delay in the
mixing is assumed. The long-dashed line shows the result for
$m_{\rm max}=70~\Msun$ and $\tau_{\rm mix}=0.1$ Gyr.}
\label{fig:mgfe_delay}
\end{figure}

We now show the influence of the different mixing timescales on the
observational constraints discussed in the previous subsections. The values of
the parameters in table~\ref{tab:parameters} are not changed.
\paragraph{[Mg/Fe] in metal-poor stars}
Figure~\ref{fig:mgfe_delay} shows the results for mixing 
time scales of 0.01, 0.1, and 1 Gyr in the [Mg/Fe]-[Fe/H] diagram. 
Since the delay due to the mixing processes elongates the timescales (see upper
x-axis for the case $\tau_{\rm mix}=0.1$ Gyr), the curve becomes flatter. 
The effect is maximum at early epochs and becomes negligible at solar ages.

The figure additionally shows the results for the inclusion of the
enrichment due to $70~\Msun$-stars and delayed mixing (TNH96-yields, 
$\tau_{\rm mix}=0.1$ Gyr, long-dashed line).
\paragraph{AMR}
While the fit to the data in the [Mg/Fe]-[Fe/H] diagram has become better,
the constraint on the AMR relation is still fulfilled. Even the results for 
mixing timescales of the order $10^{9}$ Gyr are still in agreement with
observations.
\begin{figure}
\psfig{figure=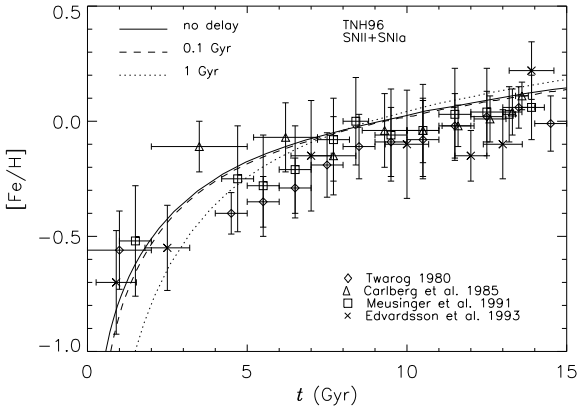,width=\linewidth}
\caption{The age-metallicity relation (AMR) for the solar neighbourhood.
For a detailed description see figure~\ref{fig:amr}. The different lines
show the influence of the different mixing timescales on the AMR relation.
For timescales of the order 0.1 Gyr, the delayed mixing only affects the
results at small $t$. In particular, the reproduction of the solar element
abundances at $t=10.45$ Gyr is not violated.}
\label{fig:amr_mix}
\end{figure}

\paragraph{Solar element abundances}
Table~\ref{tab:const_delay} gives the element abundances in both gas phases
for different mixing timescales. Since the sun forms out of {\em active} gas
at $t=10.45$ Gyr, these abundances have to match the solar values given by
observation (table~\ref{tab:constraints}). 
The abundances in the {\em inactive} gas are systematically
higher. The numbers show that the abundances of the elements H, O,
and $Z$ are well reproduced for the same set of parameters given in
table~\ref{tab:parameters}, especially for $\tau_{\rm mix}\leq 0.1$ Gyr
\begin{table}
\caption{Same as table~\ref{tab:constraints} for the two different gas
phases and various mixing timescales. Stellar yields are taken from TNH96.
The gas fractions
give the fractions of {\em active} and {\em inactive} gas to the total mass
at $t=15$ Gyr. The element abundances of the {\em active} and {\em
inactive} gas phases are given for $t=10.45$ Gyr (birth of the sun). For a
comparison with solar element abundances, the {\em active} gas phase has to
be considered. The abundances in the {\em inactive} gas are systematically
higher.}
\begin{tabular}{llll}
\hline
$\tau_{\rm mix}$      & 0.01 Gyr   & 0.1 Gyr    & 1 Gyr      \\\\
$M_g^{\rm active}/M_{\rm tot}(t_{\rm now})$   
                      & $1.30(-1)$ & $1.30(-1)$ & $1.35(-1)$  \\
$M_g^{\rm inactive}/M_{\rm tot}(t_{\rm now})$ 
                      & $6.55(-5)$ & $6.70(-4)$ & $8.38(-3)$  \\
\\
$Z^{\rm active}$      & $1.88(-2)$ & $1.91(-2)$ & $2.08(-2)$  \\
$Z^{\rm inactive}$    & $5.70(-2)$ & $5.73(-2)$ & $5.85(-2)$  \\
H$^{\rm active}$      & $6.98(-1)$ & $6.97(-1)$ & $6.92(-1)$  \\
H$^{\rm inactive}$    & $5.70(-1)$ & $5.70(-1)$ & $5.67(-1)$  \\
O$^{\rm active}$      & $9.43(-3)$ & $9.58(-3)$ & $1.05(-2)$  \\
O$^{\rm inactive}$    & $2.86(-2)$ & $2.88(-2)$ & $2.96(-2)$  \\
Mg$^{\rm active}$     & $4.57(-4)$ & $4.64(-4)$ & $5.17(-4)$  \\
Mg$^{\rm inactive}$   & $1.34(-3)$ & $1.35(-3)$ & $1.41(-3)$  \\
Fe$^{\rm active}$     & $1.23(-3)$ & $1.25(-3)$ & $1.32(-3)$  \\
Fe$^{\rm inactive}$   & $3.93(-3)$ & $3.93(-3)$ & $3.87(-3)$  \\
\hline
\end{tabular}
\label{tab:const_delay}
\end{table}

Since for a larger delay in the mixing, at the end less gas is formed into stars,
the fractions of both gas phases increase. However, more metal-poor stars
are formed in the beginning. As a consequence, the element abundances 
at $t\gg\tau_{\rm mix}$ (i.e.~when the sun is born) become higher with
increasing $\tau_{\rm mix}$. At $t\approx\tau_{\rm mix}$, instead, a larger 
delay causes lower abundances in the ISM. For the case of the iron abundance, 
this pattern is demonstrated in figure~\ref{fig:amr_mix}.

\begin{figure}
\psfig{figure=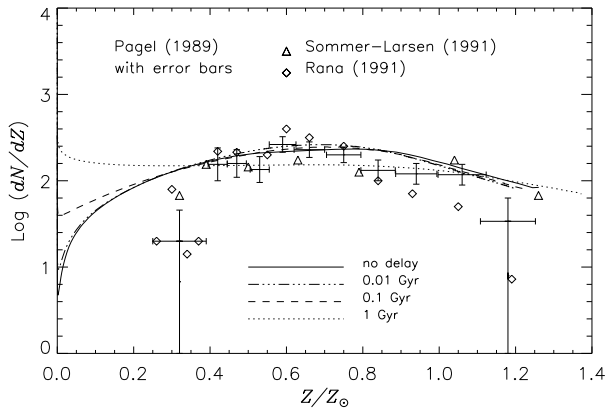,width=\linewidth}
\caption{The abundance distribution function (ADF) giving the number of stars
that are born per unit metallicity $\log (dN/dZ)$ as a function of metallicity.
For a detailed descriptions see figure~\ref{fig:adf}. 
The set of parameters is given in table~\ref{tab:constraints}. The different
linestyles show the results for different mixing timescales. The larger the
delay in the mixing processes, the more metal-poor stars form. The agreement
with the observational data becomes worse. The consideration of
pre-enrichment may be necessary.}
\label{fig:adf_mix}
\end{figure}

\paragraph{ADF}
The formation of more low-metallicity stars, though, has consequences for
the derived ADF.
The discussion in section~\ref{sec:adf} showed that the adopted infall-model
cannot fit the ADF in the whole metallicity range. At both ends of low and
high metallicity, too many stars are formed. Certainly, the inclusion of
delayed mixing worsens the situation. Figure~\ref{fig:adf_mix} shows the results
for the different mixing timescales.

There is no doubt, that the consideration of delayed mixing processes in the
disk gives a more realistic approach to the chemical evolution in the solar
neighbourhood \cite{S63,T75}. 
Since the formation of more metal-poor stars can hardly be
avoided with a delayed mixing, the additional consideration of pre-enrichment 
of the disk due to early halo evolution (Burkert, Truran, \& Hensler 1992)\nocite{BTH92}
is necessary to solve the G-dwarf problem. However, to treat this scenario 
properly, more sophisticated evolution models, calculating halo and disk 
evolution seperately, have to be considered.


\section{Conclusion}
\label{sec:conclusion}
Using two different sets of models for SN~II yields (WW95 and TNH96), we
analysed the influence of stellar nucleosynthesis on the chemical evolution
of galaxies, in particular the element abundances in the solar
neighbourhood. 

It turns out that there is a good agreement in the SN~II yields of oxygen and
total metallicity between WW95 and TNH96 over the whole mass range of SN~II.
However, from the point of view of galactic chemical evolution, there are
significant differences in the magnesium yields in the mass range
$18-25~\Msun$. For a $20~\Msun$-star, the Mg-yield calculated by TNH96 is
$\sim 5$ times higher than the result of WW95. We showed that, since the IMF
is giving more weight to smaller masses, the results of chemical evolution
models are very sensitive to this discrepancy.
The iron yield, instead, is mainly uncertain in the upper mass range.
WW95 and TNH96 agree very well in the lower mass range
($13~\Msun\leq m\leq 20~\Msun$) which is well constrained by the observed
lightcurves of SN~II events (SN1987A, SN1991J).
However, in high mass stars with $m\geq 25~\Msun$, WW95 models give
significantly higher Fe-yields than TNH96. In total, this leads to lower
[Mg/Fe]-ratios produced by WW95-nucleosynthesis. A significantly super-solar
value is only reached in high mass stars (figure~\ref{cont_mgfe}) which are
dominating the enrichment in the first few $10^7$ yr of chemical evolution.

\smallskip
Only 0.04 Gyr after the birth of the first stars, the complete generation of
SN~II-exploding stars in the mass range $8-40~\Msun$ is enriching the ISM.
We calculated the SN~II SSP-yields of O, Mg, and Fe for different IMF slopes and
both nucleosynthesis prescriptions.
The result is that TNH96-nucleosynthesis leads to ${\rm [Mg/Fe]}=0.26$ dex for
Salpeter-IMF, while the ratio with WW95 ($\Zsun$, model B) is 0.05 dex.
We showed that this discrepancy is due to a lower Mg-SSP-yield of 0.13 dex
and a higher Fe-SSP-yield of 0.08 dex in WW95. Even for a flat IMF with
$x=0.7$, the SSP-value of [Mg/Fe] is 0.12 dex with WW95-yields. Without any
impact from complex evolution models, from these
numbers one can already conclude, that the [Mg/Fe] overabundance in both,
ellipticals and the solar neighbourhood, cannot be explained with the stellar 
yields of the WW95-models.

\smallskip
Applying the standard infall-model \cite{MG86,TWW95,YTN96} on the chemical 
evolution of the solar neighbourhood confirms the conclusions drawn from the
discussion of the SSP yields.
Both, the [Mg/Fe] overabundance in metal-poor stars
and the magnesium abundance of the sun can be better reproduced with the
Mg-yields of TNH96.
In addition to this, we discussed the relaxation of the instantaneous
{\em mixing} approximation for the chemical evolution in the solar
neighbourhood. For this purpose, we modified the basic equations of chemical
evolution seperating the gaseous component into two different gas phases.
While the {\em inactive} phase is enriched by the stellar ejecta, stars can
only form out of the {\em active}, well mixed phase. A mass flow from the
{\em inactive} to the {\em active} gas phase on a variable timescale
represents the mixing process. For different mixing timescales of the order
$10^7,10^8,10^9$ yr, we investigated the influence of a delayed mixing on the 
reproduction of the observational features. It turns out that 
a delay in the mixing supports the approximately constant value of
[Mg/Fe] in the [Mg/Fe]-[Fe/H]-diagram in the low-metallicity range, while
the agreement with the age-metallicity relation (AMR) and solar element 
abundances is not violated. However, since a delayed mixing causes the
formation of more low-metallicity stars, the abundance distribution 
function (ADF) is less well reproduced. On the other hand, the instantaneous 
mixing of the stellar ejecta is certainly an unrealistic assumption, and the 
inclusion of a delay is a necessary step to improve chemical evolution models.
Hence, the solution of the G-dwarf problem in the solar neighbourhood may 
require a combination of infall and pre-enrichment.

\smallskip
Since TNH96 include the $70~\Msun$-star in their computations, we
additionally investigated the influence of a variation of the upper mass
cutoff on the theoretical SSP-yields and on the chemical evolution in the
solar neighbourhood. Applying TNH96 nucleosynthesis, the [Mg/Fe] ratio in
the ejecta of one SSP is significantly increased. This result is highly
uncertain, however, because TNH96 do not consider fall back, which may play
an important role for the nucleosynthetic contribution from high mass stars.
Indeed, extrapolating the results of WW95 to $70~\Msun$ leaves the SSP yields
basically unchanged. The problem of the underestimation of the solar
magnesium abundance remains the same also for TNH96-yields, since the Mg/O
ratio in stars more massive than $40~\Msun$ is even smaller. However, it
is important to investigate quantitatively the metal contribution of stars
more massive than $40~\Msun$, since they could play an important role for
chemical evolution.

\smallskip
In general, we demonstrated the sensitivity of galactic chemical evolution on
nucleosynthesis prescriptions of type II supernovae. Different stellar
yields can significantly alter conclusions on the parameters of chemical
evolution models like IMF slope or star formation timescales. As long as the
stellar nucleosynthesis of important elements like magnesium and iron is
affected by so many uncertainties, the results from simulations of chemical
evolution have to be interpreted considering
the whole range of up to date nucleosynthesis calculations.

\section{Acknowledgements}
We would like to thank B.~Pagel, the referee of the paper, for carefully
reading the first version and giving important remarks on the subject. He
especially motivated us to explore the influence of the upper mass cutoff on
the results of the calculations. We also thank F.-K. Thielemann for usefull
and interesting discussions.
The SFB 375 of the DFG and the Alexander von Humboldt Stiftung (L.G.) are 
acknowledged for support.




\appendix
\section{IMF weighted yields}
\label{ap:cont}
The figures show the relative contribution of a $1~\Msun$ mass interval to the
total SN~II yields of the elements oxygen, magnesium, iron, and metallicity for
different IMF slopes and nucleosynthesis prescriptions (WW95, TNH96). The
three models of different explosion energies (A,B,C)\footnote{See also
table~\ref{tab:WW_TNH}.} and four different initial 
metallicities ($10^{-4}\Zsun,0.01\Zsun,0.1\Zsun,\Zsun$) in WW95 are considered.
Each figure shows the results for one certain element and one specified
WW95-model (A,B,C). The four panels of each figure show the results for the
four different IMF slopes. The four different initial metallicities of WW95
and TNH96-yields are plotted together in each panel. The meaning of
the linestyles and symbols are specified in figure~\ref{eject_hydro}.
The quantity $dQ_{im}/dm$ is determined according to
equation~\ref{eq:cont}. It is normalized such that the integration
over the total mass range of SN~II ($11-40~\Msun$) is equal 1.
The x-axis give the initial stellar mass on the main sequence. To obtain the
fractional contribution of a mass interval, one has to multiply the width
of the interval ($\Msun$) with the mean value of $dQ_{im}/dm$ in this mass
range.

\subsection{Oxygen}
\begin{figure}
\psfig{figure=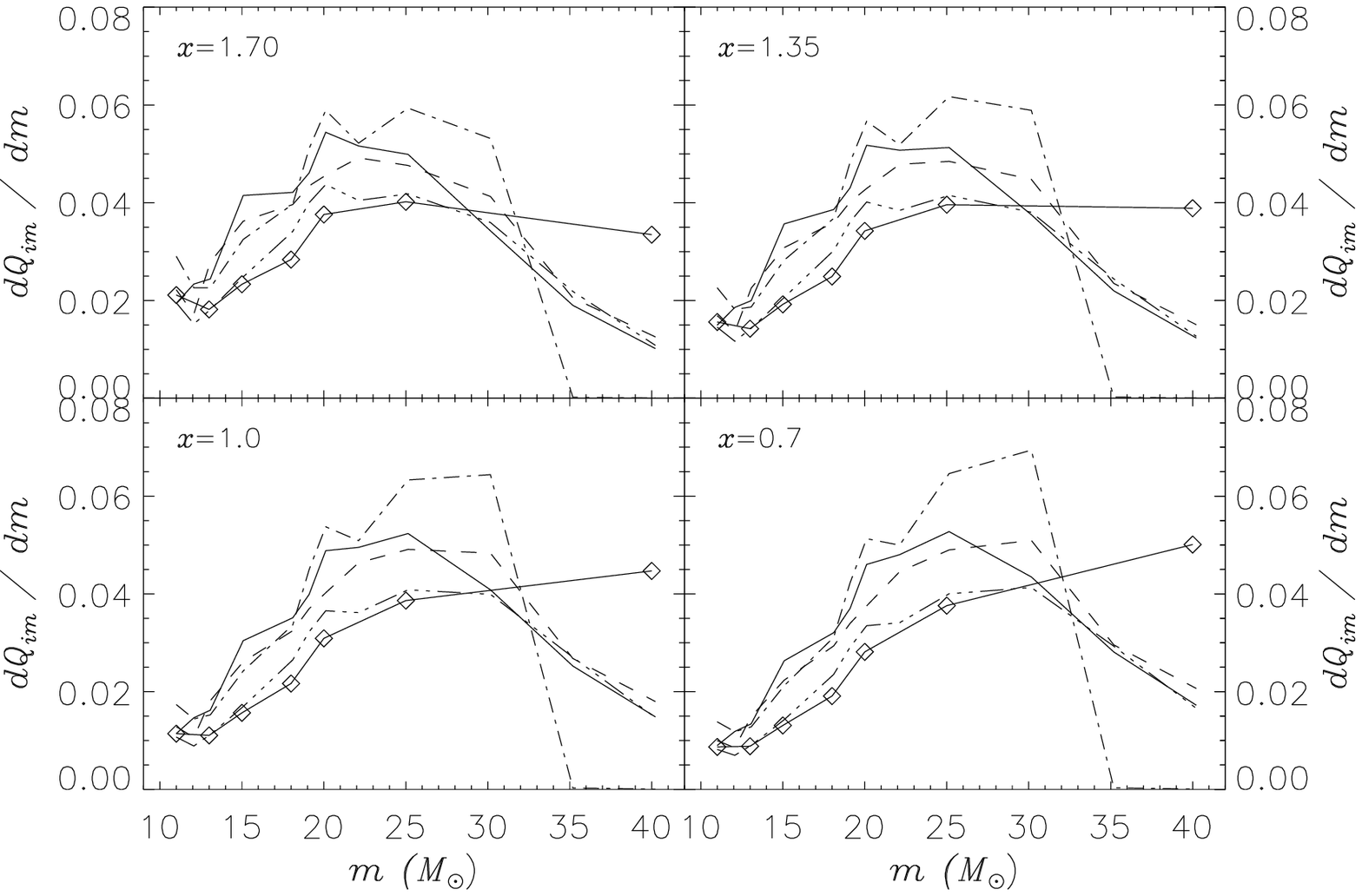,width=\linewidth}
\caption{Oxygen. Model A.}
\label{cont_oxy_a}
\end{figure}
\begin{figure}
\psfig{figure=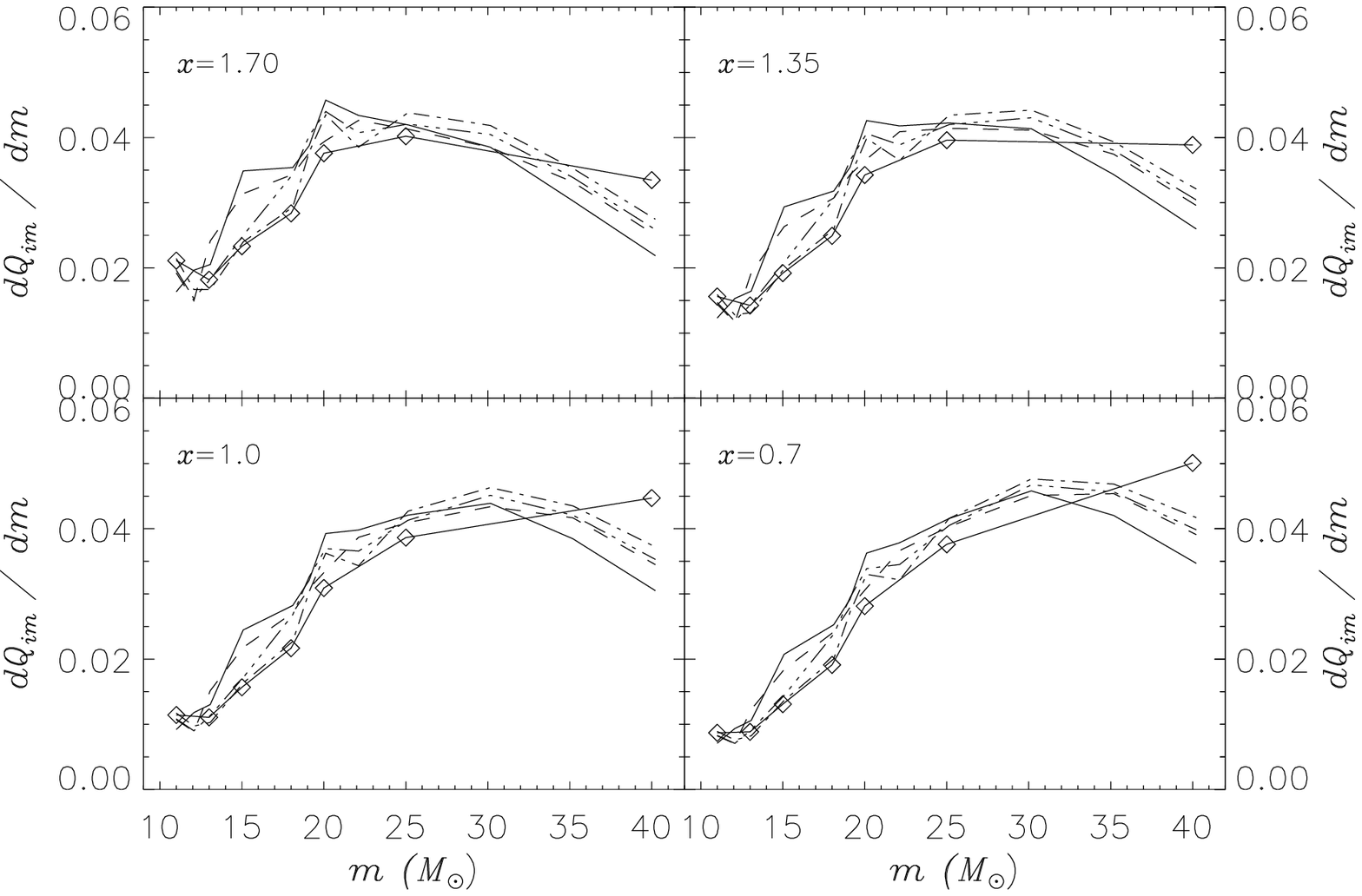,width=\linewidth}
\caption{Oxygen. Model B.}
\label{cont_oxy_b}
\end{figure}
\begin{figure}
\psfig{figure=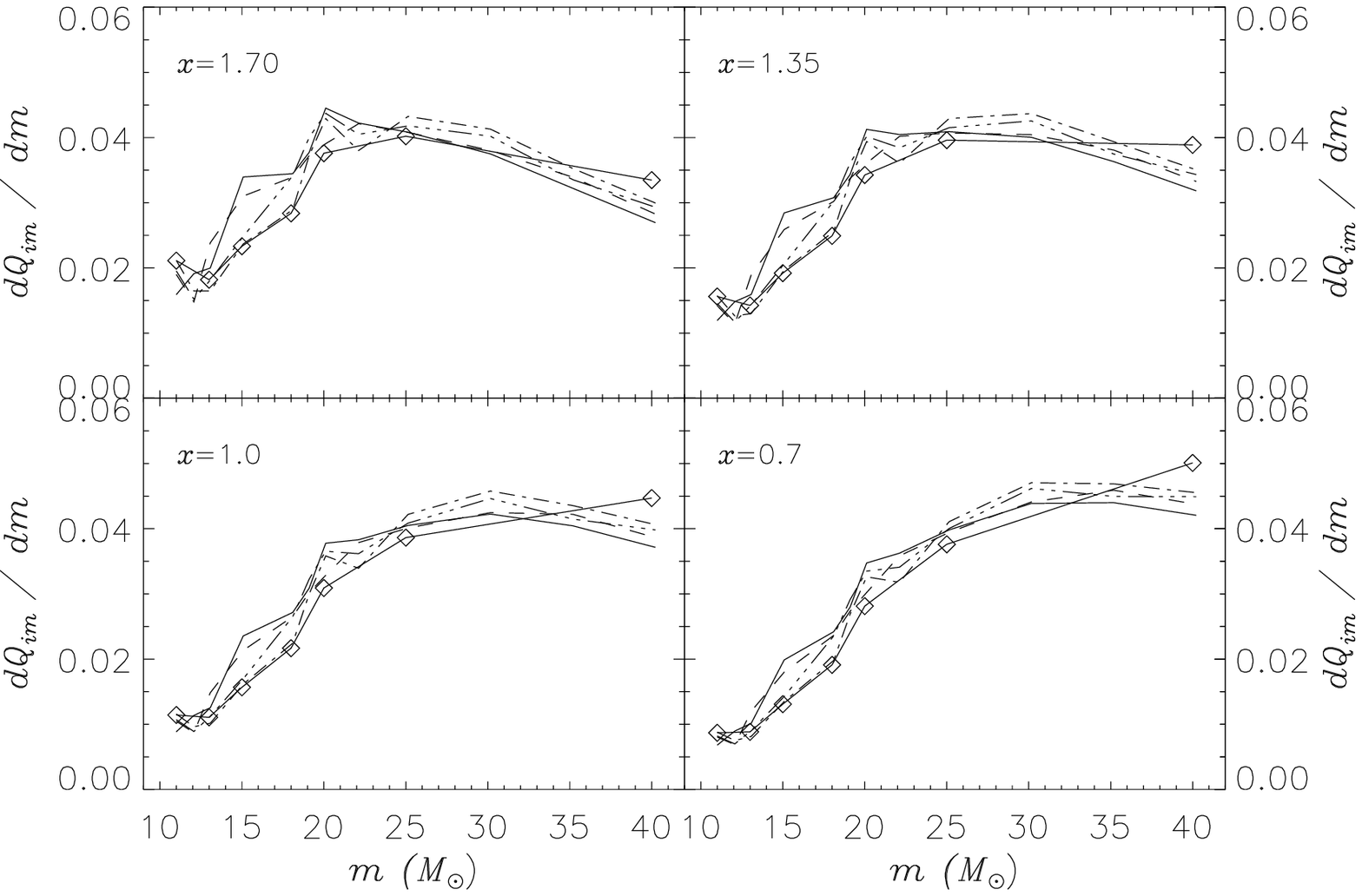,width=\linewidth}
\caption{Oxygen. Model C.}
\label{cont_oxy_c}
\end{figure}

\subsection{Metallicity}
\begin{figure}
\psfig{figure=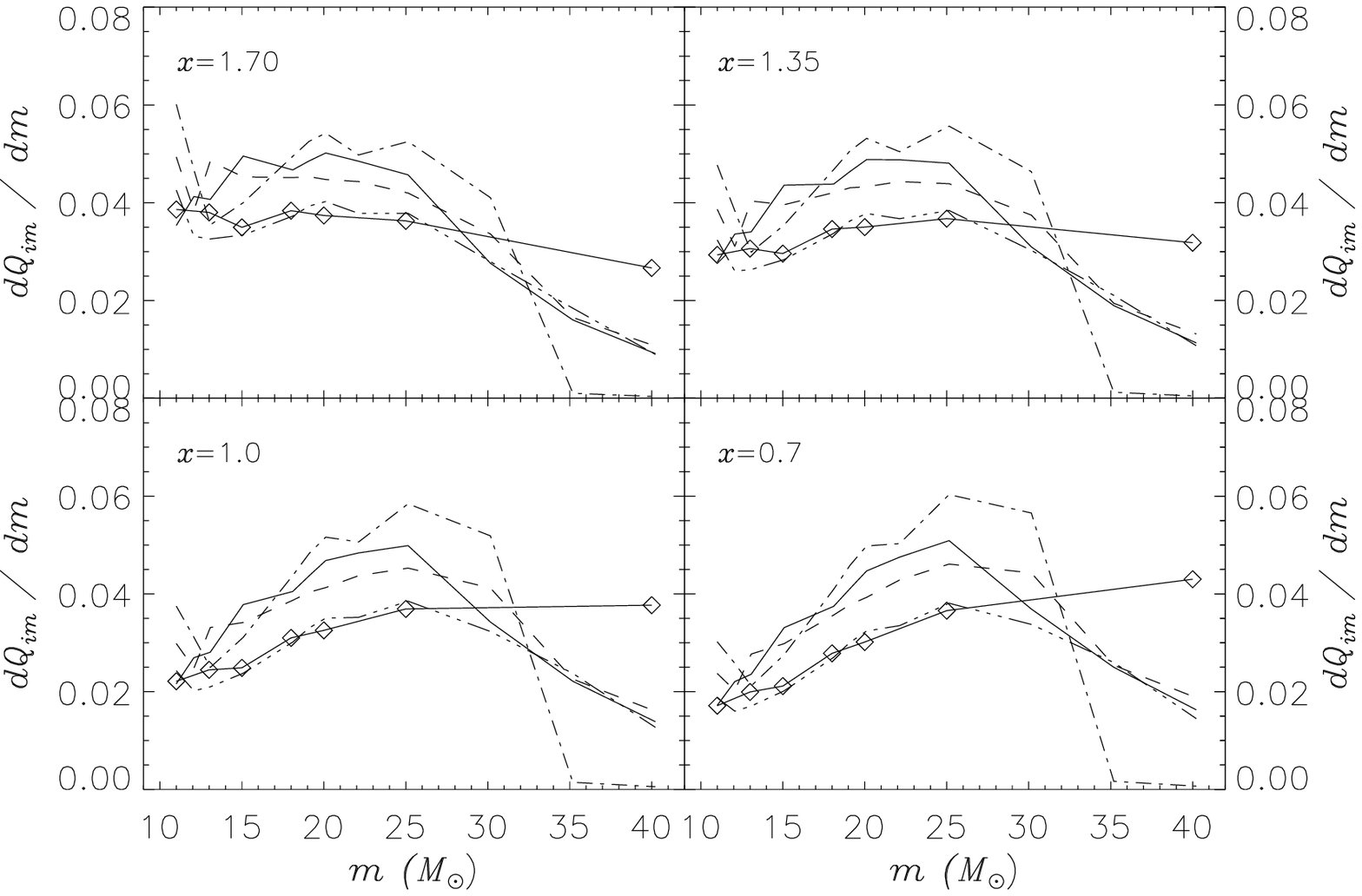,width=\linewidth}
\caption{Metallicity. Model A.}
\label{cont_z_a}
\end{figure}
\begin{figure}
\psfig{figure=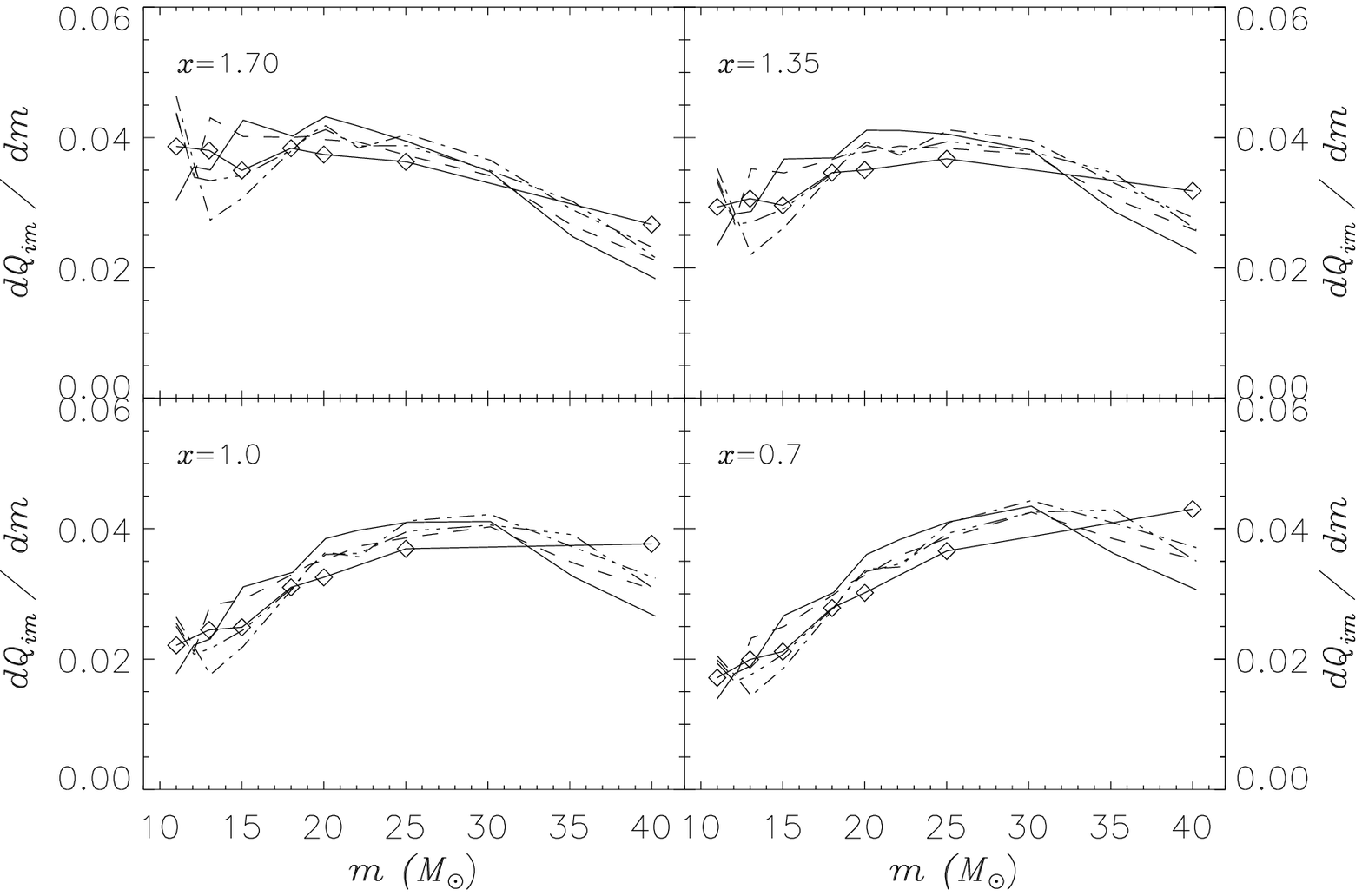,width=\linewidth}
\caption{Metallicity. Model B.}
\label{cont_z_b}
\end{figure}
\begin{figure}
\psfig{figure=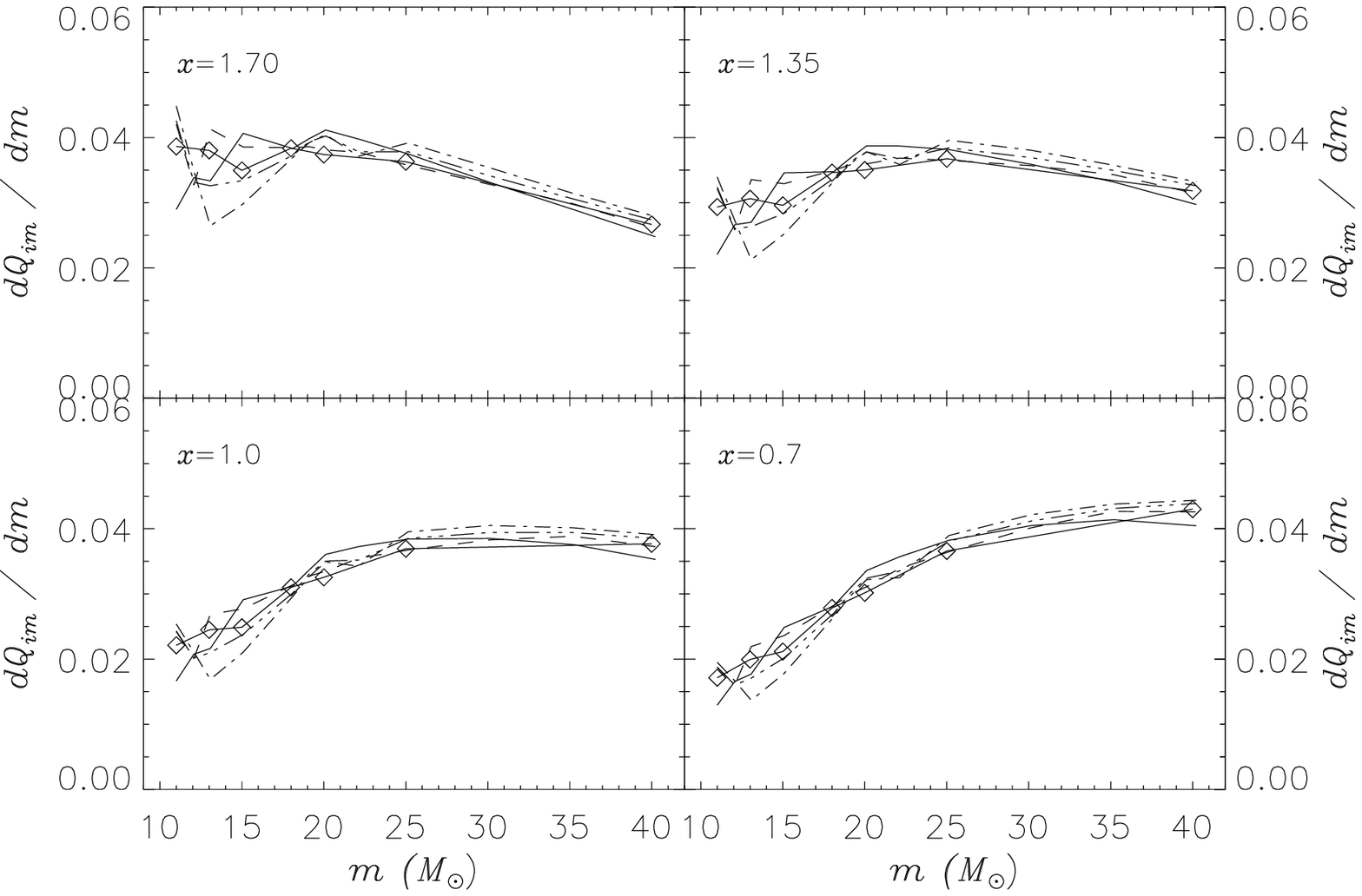,width=\linewidth}
\caption{Metallicity. Model C.}
\label{cont_z_c}
\end{figure}

\subsection{Magnesium}
\begin{figure}
\psfig{figure=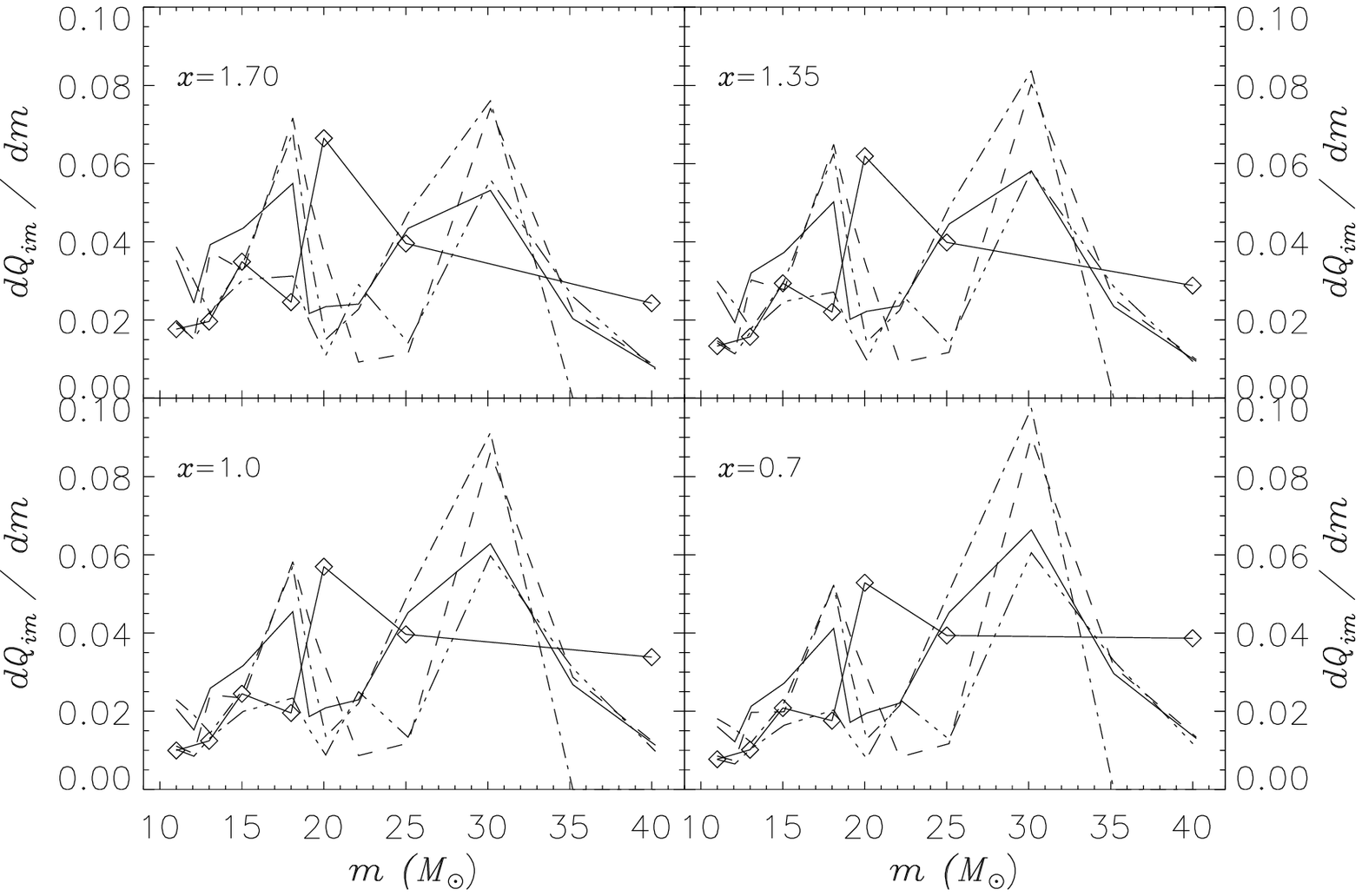,width=\linewidth}
\caption{Magnesium. Model A.}
\label{cont_mag_a}
\end{figure}
\begin{figure}
\psfig{figure=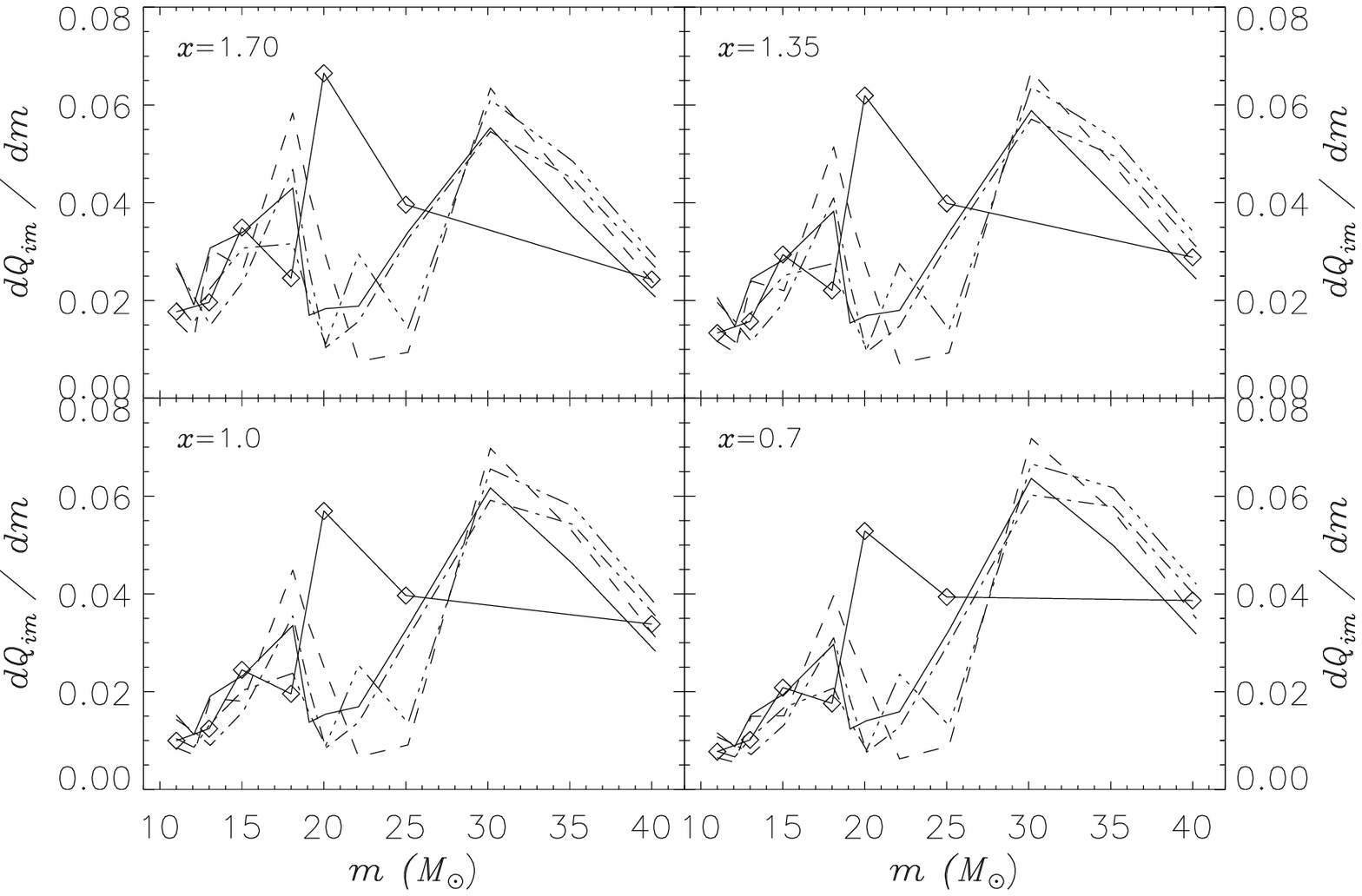,width=\linewidth}
\caption{Magnesium. Model B.}
\label{cont_mag_b}
\end{figure}
\begin{figure}
\psfig{figure=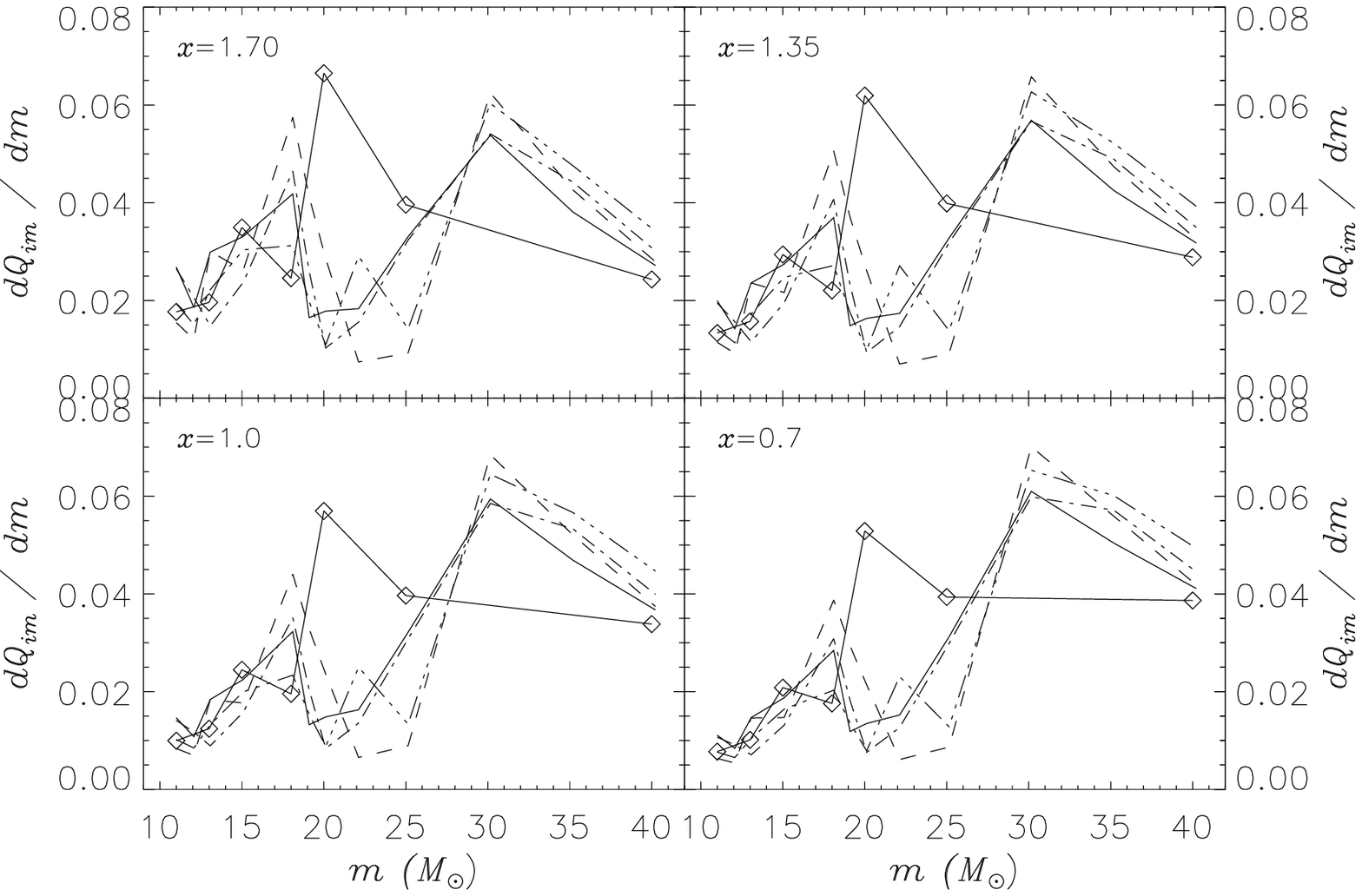,width=\linewidth}
\caption{Magnesium. Model C.}
\label{cont_mag_c}
\end{figure}

\subsection{Iron}
\begin{figure}
\psfig{figure=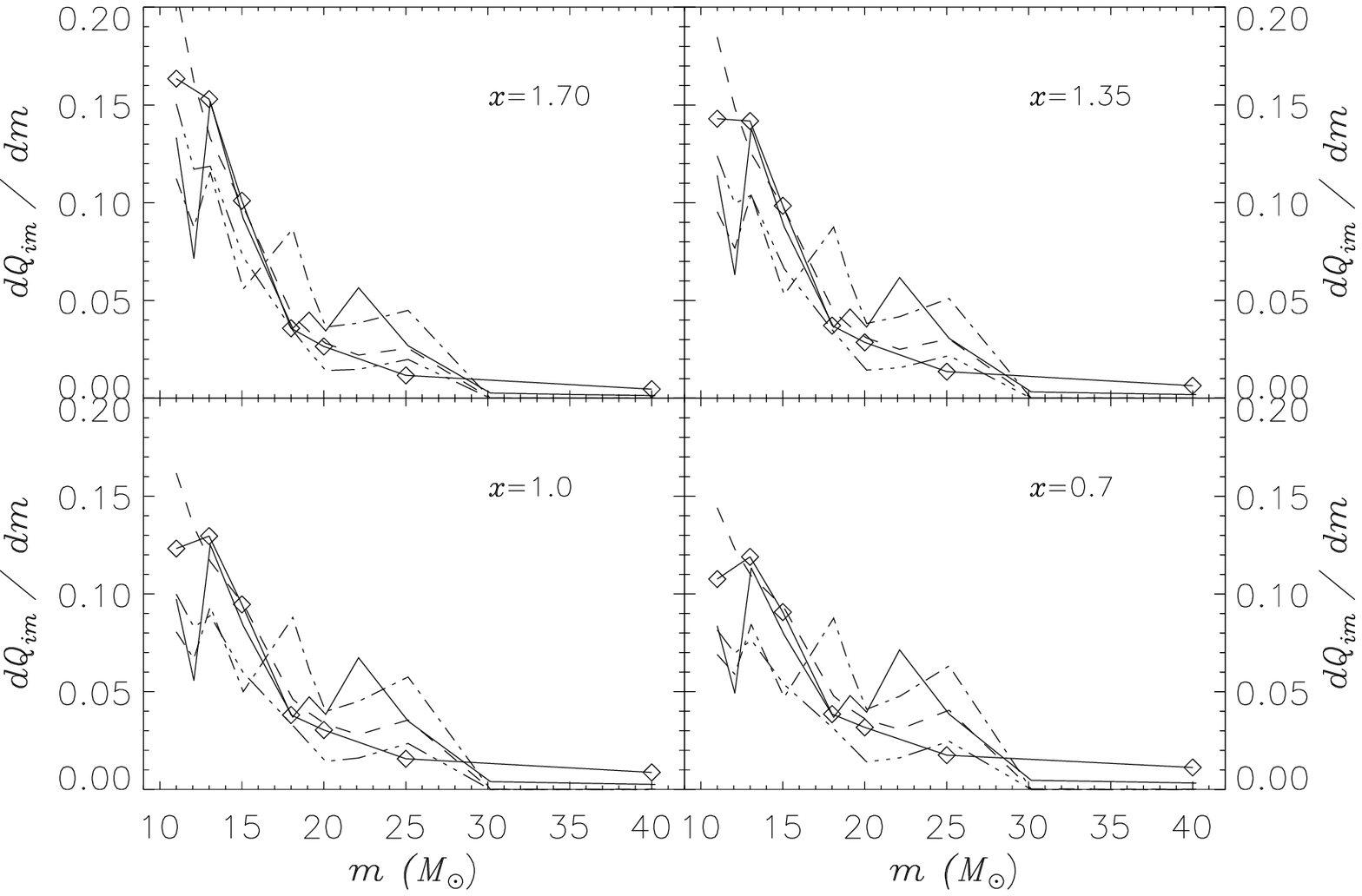,width=\linewidth}
\caption{Iron. Model A.}
\label{cont_fe_a}
\end{figure}
\begin{figure}
\psfig{figure=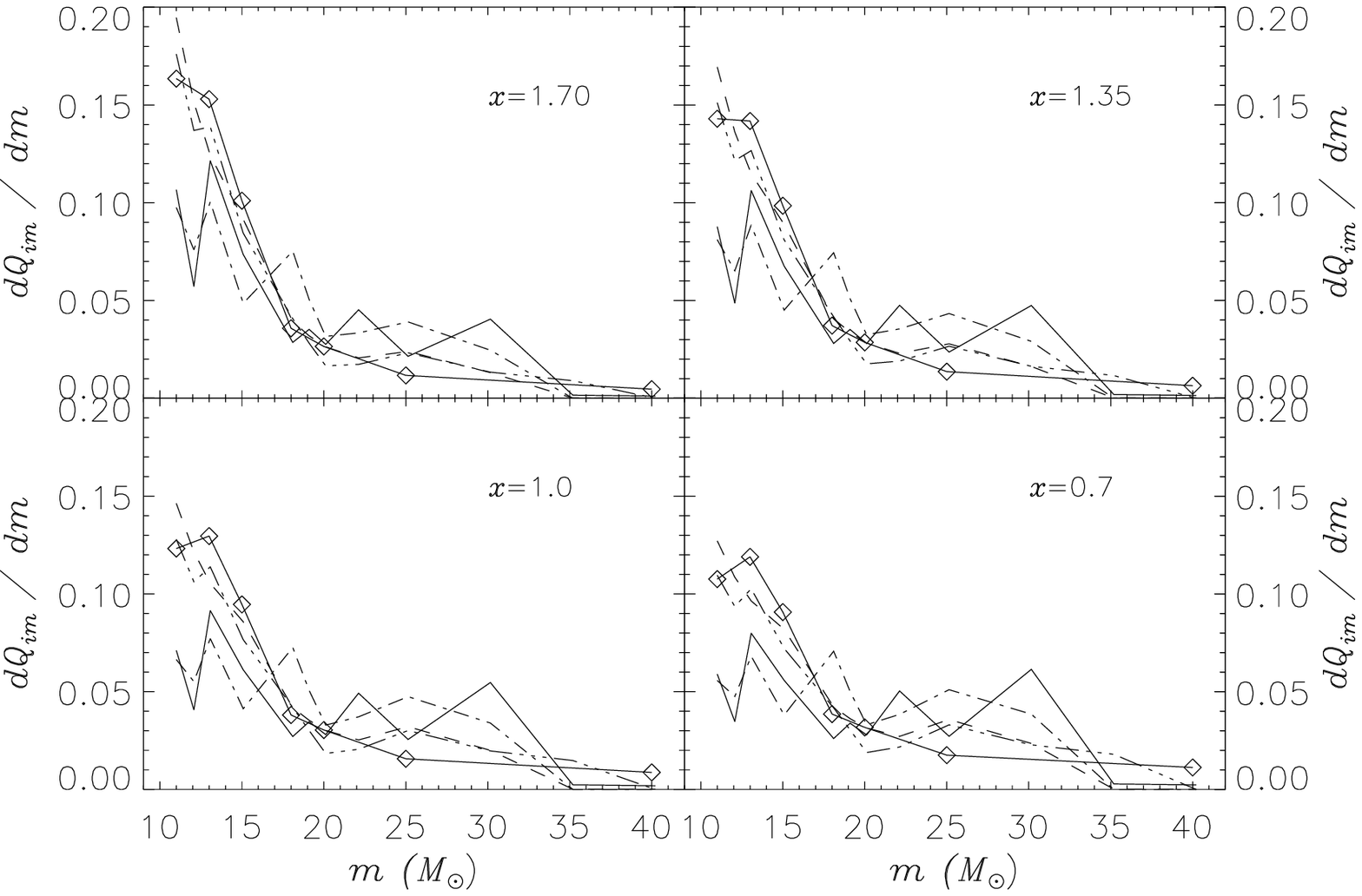,width=\linewidth}
\caption{Iron. Model B.}
\label{cont_fe_b}
\end{figure}
\begin{figure}
\psfig{figure=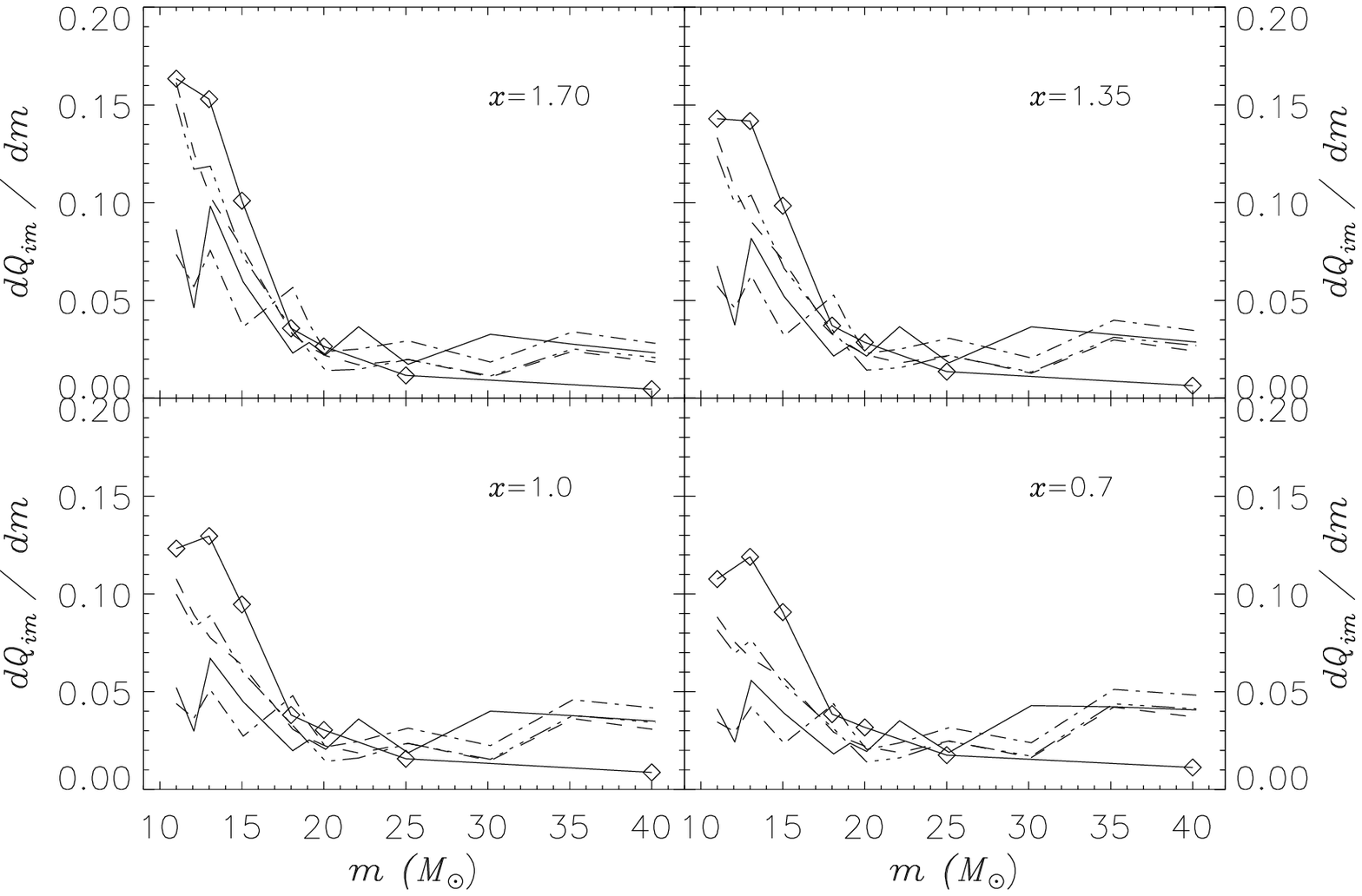,width=\linewidth}
\caption{Iron. Model C.}
\label{cont_fe_c}
\end{figure}

\clearpage

\section{SSP-yields}
\label{ap:tables}
The tables~\ref{tab:A1.70} to \ref{tab:C0.70}
show the abundances of oxygen, magnesium and iron in the ejecta of 
one generation of SN~II-exploding stars (SSP-yields). One table refers to a
certain IMF-slope and explosion model of WW95 (A,B,C). The values are
normalized on (meteoritic) solar abundances from Anders \& Grevesse
\shortcite{AG89} and expressed on a logarithmic scale.

\subsection{$\bmath{x=1.7}$}
\begin{table*}
\begin{minipage}{10.5cm}
\caption{WW95-model: A. IMF: $x=1.70$.}
\begin{tabular}{rrrrrrr}
\hline
           &   TNH96 &    WW95 &    WW95 &    WW95 &    WW95 &    WW95\\
& & $(0\Zsun)$ & $(10^{-4}\Zsun)$ & $(0.01\Zsun$) & $(0.1\Zsun$) & $(\Zsun)$\\\\
     {[O]} & $ 0.85$ & $ 0.04$ & $ 0.61$ & $ 0.72$ & $ 0.73$ & $ 0.76$\\
    {[Mg]} & $ 0.85$ & $ 0.12$ & $ 0.52$ & $ 0.56$ & $ 0.64$ & $ 0.60$\\
    {[Fe]} & $ 0.65$ & $ 0.52$ & $ 0.51$ & $ 0.71$ & $ 0.76$ & $ 0.61$\\
  {[O/Fe]} & $ 0.19$ & $-0.48$ & $ 0.10$ & $ 0.00$ & $-0.03$ & $ 0.15$\\
 {[Mg/Fe]} & $ 0.19$ & $-0.40$ & $ 0.01$ & $-0.15$ & $-0.12$ & $-0.01$\\
\hline
\end{tabular}
\label{tab:A1.70}
\end{minipage}
\end{table*}

\begin{table*}
\begin{minipage}{10.5cm}
\caption{WW95-model: B. IMF: $x=1.70$.}
\begin{tabular}{rrrrrrr}
\hline
           &   TNH96 &    WW95 &    WW95 &    WW95 &    WW95 &    WW95\\
& & $(0\Zsun)$ & $(10^{-4}\Zsun)$ & $(0.01\Zsun$) & $(0.1\Zsun$) & $(\Zsun)$\\\\
     {[O]} & $ 0.85$ & $ 0.43$ & $ 0.75$ & $ 0.77$ & $ 0.79$ & $ 0.83$\\
    {[Mg]} & $ 0.85$ & $ 0.44$ & $ 0.68$ & $ 0.65$ & $ 0.73$ & $ 0.71$\\
    {[Fe]} & $ 0.65$ & $ 0.68$ & $ 0.57$ & $ 0.76$ & $ 0.79$ & $ 0.70$\\
  {[O/Fe]} & $ 0.19$ & $-0.26$ & $ 0.17$ & $ 0.01$ & $ 0.00$ & $ 0.13$\\
 {[Mg/Fe]} & $ 0.19$ & $-0.25$ & $ 0.11$ & $-0.11$ & $-0.06$ & $ 0.00$\\
\hline
\end{tabular}
\label{tab:B1.70}
\end{minipage}
\end{table*}

\begin{table*}
\begin{minipage}{10.5cm}
\caption{WW95-model: C. IMF: $x=1.70$.}
\begin{tabular}{rrrrrrr}
\hline
           &   TNH96 &    WW95 &    WW95 &    WW95 &    WW95 &    WW95\\
& & $(0\Zsun)$ & $(10^{-4}\Zsun)$ & $(0.01\Zsun$) & $(0.1\Zsun$) & $(\Zsun)$\\\\
     {[O]} & $ 0.85$ & $ 0.66$ & $ 0.75$ & $ 0.78$ & $ 0.80$ & $ 0.85$\\
    {[Mg]} & $ 0.85$ & $ 0.63$ & $ 0.69$ & $ 0.66$ & $ 0.73$ & $ 0.72$\\
    {[Fe]} & $ 0.65$ & $ 0.78$ & $ 0.70$ & $ 0.83$ & $ 0.87$ & $ 0.80$\\
  {[O/Fe]} & $ 0.19$ & $-0.12$ & $ 0.06$ & $-0.06$ & $-0.07$ & $ 0.05$\\
 {[Mg/Fe]} & $ 0.19$ & $-0.15$ & $-0.01$ & $-0.17$ & $-0.14$ & $-0.08$\\
\hline
\end{tabular}
\label{tab:C1.70}
\end{minipage}
\end{table*}

\subsection{$\bmath{x=1.35}$}
\begin{table*}
\begin{minipage}{10.5cm}
\caption{WW95-model: A. IMF: $x=1.35$.}
\begin{tabular}{rrrrrrr}
\hline
           &   TNH96 &    WW95 &    WW95 &    WW95 &    WW95 &    WW95\\
& & $(0\Zsun)$ & $(10^{-4}\Zsun)$ & $(0.01\Zsun$) & $(0.1\Zsun$) & $(\Zsun)$\\\\
     {[O]} & $ 0.92$ & $ 0.05$ & $ 0.66$ & $ 0.78$ & $ 0.79$ & $ 0.81$\\
    {[Mg]} & $ 0.91$ & $ 0.13$ & $ 0.57$ & $ 0.63$ & $ 0.69$ & $ 0.66$\\
    {[Fe]} & $ 0.65$ & $ 0.50$ & $ 0.52$ & $ 0.70$ & $ 0.75$ & $ 0.62$\\
  {[O/Fe]} & $ 0.26$ & $-0.45$ & $ 0.14$ & $ 0.07$ & $ 0.04$ & $ 0.20$\\
 {[Mg/Fe]} & $ 0.26$ & $-0.37$ & $ 0.05$ & $-0.08$ & $-0.05$ & $ 0.04$\\
\hline
\end{tabular}
\label{tab:A1.35}
\end{minipage}
\end{table*}

\begin{table*}
\begin{minipage}{10.5cm}
\caption{WW95-model: B. IMF: $x=1.35$.}
\begin{tabular}{rrrrrrr}
\hline
           &   TNH96 &    WW95 &    WW95 &    WW95 &    WW95 &    WW95\\
& & $(0\Zsun)$ & $(10^{-4}\Zsun)$ & $(0.01\Zsun$) & $(0.1\Zsun$) & $(\Zsun)$\\\\
     {[O]} & $ 0.92$ & $ 0.49$ & $ 0.82$ & $ 0.84$ & $ 0.86$ & $ 0.90$\\
    {[Mg]} & $ 0.91$ & $ 0.49$ & $ 0.76$ & $ 0.73$ & $ 0.80$ & $ 0.78$\\
    {[Fe]} & $ 0.65$ & $ 0.70$ & $ 0.59$ & $ 0.77$ & $ 0.79$ & $ 0.73$\\
  {[O/Fe]} & $ 0.26$ & $-0.21$ & $ 0.22$ & $ 0.07$ & $ 0.07$ & $ 0.17$\\
 {[Mg/Fe]} & $ 0.26$ & $-0.21$ & $ 0.16$ & $-0.04$ & $ 0.01$ & $ 0.05$\\
\hline
\end{tabular}
\label{tab:B1.35}
\end{minipage}
\end{table*}

\begin{table*}
\begin{minipage}{10.5cm}
\caption{WW95-model: C. IMF: $x=1.35$.}
\begin{tabular}{rrrrrrr}
\hline
           &   TNH96 &    WW95 &    WW95 &    WW95 &    WW95 &    WW95\\
& & $(0\Zsun)$ & $(10^{-4}\Zsun)$ & $(0.01\Zsun$) & $(0.1\Zsun$) & $(\Zsun)$\\\\
     {[O]} & $ 0.92$ & $ 0.74$ & $ 0.82$ & $ 0.84$ & $ 0.86$ & $ 0.91$\\
    {[Mg]} & $ 0.91$ & $ 0.70$ & $ 0.76$ & $ 0.74$ & $ 0.80$ & $ 0.79$\\
    {[Fe]} & $ 0.65$ & $ 0.82$ & $ 0.74$ & $ 0.86$ & $ 0.89$ & $ 0.84$\\
  {[O/Fe]} & $ 0.26$ & $-0.08$ & $ 0.08$ & $-0.01$ & $-0.03$ & $ 0.07$\\
 {[Mg/Fe]} & $ 0.26$ & $-0.12$ & $ 0.02$ & $-0.12$ & $-0.09$ & $-0.05$\\
\hline
\end{tabular}
\label{tab:C1.35}
\end{minipage}
\end{table*}

\subsection{$\bmath{x=1.0}$}
\begin{table*}
\begin{minipage}{10.5cm}
\caption{WW95-model: A. IMF: $x=1.00$.}
\begin{tabular}{rrrrrrr}
\hline
           &   TNH96 &    WW95 &    WW95 &    WW95 &    WW95 &    WW95\\
& & $(0\Zsun)$ & $(10^{-4}\Zsun)$ & $(0.01\Zsun$) & $(0.1\Zsun$) & $(\Zsun)$\\\\
     {[O]} & $ 0.98$ & $ 0.07$ & $ 0.71$ & $ 0.83$ & $ 0.84$ & $ 0.86$\\
    {[Mg]} & $ 0.97$ & $ 0.13$ & $ 0.62$ & $ 0.69$ & $ 0.75$ & $ 0.71$\\
    {[Fe]} & $ 0.65$ & $ 0.47$ & $ 0.53$ & $ 0.69$ & $ 0.74$ & $ 0.62$\\
  {[O/Fe]} & $ 0.33$ & $-0.41$ & $ 0.18$ & $ 0.14$ & $ 0.10$ & $ 0.24$\\
 {[Mg/Fe]} & $ 0.32$ & $-0.35$ & $ 0.09$ & $ 0.00$ & $ 0.01$ & $ 0.09$\\
\hline
\end{tabular}
\label{tab:A1.00}
\end{minipage}
\end{table*}

\begin{table*}
\begin{minipage}{10.5cm}
\caption{WW95-model: B. IMF: $x=1.00$.}
\begin{tabular}{rrrrrrr}
\hline
           &   TNH96 &    WW95 &    WW95 &    WW95 &    WW95 &    WW95\\
& & $(0\Zsun)$ & $(10^{-4}\Zsun)$ & $(0.01\Zsun$) & $(0.1\Zsun$) & $(\Zsun)$\\\\
     {[O]} & $ 0.98$ & $ 0.55$ & $ 0.88$ & $ 0.90$ & $ 0.92$ & $ 0.95$\\
    {[Mg]} & $ 0.97$ & $ 0.54$ & $ 0.82$ & $ 0.80$ & $ 0.86$ & $ 0.84$\\
    {[Fe]} & $ 0.65$ & $ 0.71$ & $ 0.61$ & $ 0.77$ & $ 0.78$ & $ 0.75$\\
  {[O/Fe]} & $ 0.33$ & $-0.16$ & $ 0.27$ & $ 0.13$ & $ 0.13$ & $ 0.20$\\
 {[Mg/Fe]} & $ 0.32$ & $-0.17$ & $ 0.21$ & $ 0.03$ & $ 0.08$ & $ 0.09$\\
\hline
\end{tabular}
\label{tab:B1.00}
\end{minipage}
\end{table*}

\begin{table*}
\begin{minipage}{10.5cm}
\caption{WW95-model: C. IMF: $x=1.00$.}
\begin{tabular}{rrrrrrr}
\hline
           &   TNH96 &    WW95 &    WW95 &    WW95 &    WW95 &    WW95\\
& & $(0\Zsun)$ & $(10^{-4}\Zsun)$ & $(0.01\Zsun$) & $(0.1\Zsun$) & $(\Zsun)$\\\\
     {[O]} & $ 0.98$ & $ 0.81$ & $ 0.89$ & $ 0.91$ & $ 0.92$ & $ 0.97$\\
    {[Mg]} & $ 0.97$ & $ 0.77$ & $ 0.83$ & $ 0.81$ & $ 0.87$ & $ 0.85$\\
    {[Fe]} & $ 0.65$ & $ 0.86$ & $ 0.79$ & $ 0.88$ & $ 0.91$ & $ 0.89$\\
  {[O/Fe]} & $ 0.33$ & $-0.05$ & $ 0.10$ & $ 0.03$ & $ 0.01$ & $ 0.08$\\
 {[Mg/Fe]} & $ 0.32$ & $-0.09$ & $ 0.04$ & $-0.07$ & $-0.04$ & $-0.03$\\
\hline
\end{tabular}
\label{tab:C1.00}
\end{minipage}
\end{table*}

\subsection{$\bmath{x=0.7}$}
\begin{table*}
\begin{minipage}{10.5cm}
\caption{WW95-model: A. IMF: $x=0.70$.}
\begin{tabular}{rrrrrrr}
\hline
           &   TNH96 &    WW95 &    WW95 &    WW95 &    WW95 &    WW95\\
& & $(0\Zsun)$ & $(10^{-4}\Zsun)$ & $(0.01\Zsun$) & $(0.1\Zsun$) & $(\Zsun)$\\\\
     {[O]} & $ 1.03$ & $ 0.07$ & $ 0.74$ & $ 0.87$ & $ 0.88$ & $ 0.89$\\
    {[Mg]} & $ 1.01$ & $ 0.12$ & $ 0.65$ & $ 0.73$ & $ 0.79$ & $ 0.74$\\
    {[Fe]} & $ 0.64$ & $ 0.45$ & $ 0.52$ & $ 0.67$ & $ 0.72$ & $ 0.61$\\
  {[O/Fe]} & $ 0.39$ & $-0.38$ & $ 0.21$ & $ 0.20$ & $ 0.16$ & $ 0.28$\\
 {[Mg/Fe]} & $ 0.37$ & $-0.32$ & $ 0.13$ & $ 0.06$ & $ 0.07$ & $ 0.13$\\
\hline
\end{tabular}
\label{tab:A0.70}
\end{minipage}
\end{table*}

\begin{table*}
\begin{minipage}{10.5cm}
\caption{WW95-model: B. IMF: $x=0.70$.}
\begin{tabular}{rrrrrrr}
\hline
           &   TNH96 &    WW95 &    WW95 &    WW95 &    WW95 &    WW95\\
& & $(0\Zsun)$ & $(10^{-4}\Zsun)$ & $(0.01\Zsun$) & $(0.1\Zsun$) & $(\Zsun)$\\\\
     {[O]} & $ 1.03$ & $ 0.60$ & $ 0.93$ & $ 0.95$ & $ 0.96$ & $ 1.00$\\
    {[Mg]} & $ 1.01$ & $ 0.58$ & $ 0.88$ & $ 0.86$ & $ 0.91$ & $ 0.89$\\
    {[Fe]} & $ 0.64$ & $ 0.72$ & $ 0.62$ & $ 0.77$ & $ 0.77$ & $ 0.76$\\
  {[O/Fe]} & $ 0.39$ & $-0.12$ & $ 0.31$ & $ 0.18$ & $ 0.19$ & $ 0.23$\\
 {[Mg/Fe]} & $ 0.37$ & $-0.14$ & $ 0.26$ & $ 0.09$ & $ 0.14$ & $ 0.12$\\
\hline
\end{tabular}
\label{tab:B0.70}
\end{minipage}
\end{table*}

\begin{table*}
\begin{minipage}{10.5cm}
\caption{WW95-model: C. IMF: $x=0.70$.}
\begin{tabular}{rrrrrrr}
\hline
           &   TNH96 &    WW95 &    WW95 &    WW95 &    WW95 &    WW95\\
& & $(0\Zsun)$ & $(10^{-4}\Zsun)$ & $(0.01\Zsun$) & $(0.1\Zsun$) & $(\Zsun)$\\\\
     {[O]} & $ 1.03$ & $ 0.87$ & $ 0.94$ & $ 0.95$ & $ 0.97$ & $ 1.02$\\
    {[Mg]} & $ 1.01$ & $ 0.82$ & $ 0.88$ & $ 0.86$ & $ 0.92$ & $ 0.90$\\
    {[Fe]} & $ 0.64$ & $ 0.89$ & $ 0.83$ & $ 0.90$ & $ 0.93$ & $ 0.92$\\
  {[O/Fe]} & $ 0.39$ & $-0.02$ & $ 0.11$ & $ 0.05$ & $ 0.04$ & $ 0.10$\\
 {[Mg/Fe]} & $ 0.37$ & $-0.07$ & $ 0.06$ & $-0.04$ & $-0.01$ & $-0.01$\\
\hline
\end{tabular}
\label{tab:C0.70}
\end{minipage}
\end{table*}

\bsp
\end{document}